\documentclass[showpacs,floatfix,amsmath,amssymb,pre]{revtex4}

\usepackage{latexsym,exscale}
\usepackage[ansinew]{inputenc}
\usepackage{amsmath}
\usepackage{textcomp}
\usepackage{amsfonts}
\usepackage{amssymb}
\usepackage{mathrsfs}
\usepackage{psfrag}
\usepackage{color} 
\usepackage{graphicx}
\usepackage{pstricks}
\usepackage{subfigure}
\usepackage{anysize}
\usepackage[T1]{fontenc}
\usepackage{float}
\usepackage{booktabs}
\usepackage{geometry}
\usepackage{fancyhdr}
\usepackage{bm}% bold math
\usepackage{sidecap} %bildbeschriftung seitlich

\usepackage{lmodern}						% Ersatz fuer Computer Modern-Schriften

\newcommand{\colr}[1]{\textcolor{black}{#1}}

\begin{document}

%----------------------------- HEADER ----------------------------------
\title{Onset of Synchronization in Complex Networks of Noisy Oscillators} 
\author{Bernard Sonnenschein}
\email{sonne@physik.hu-berlin.de}
\author{Lutz Schimansky-Geier}
\affiliation{Department of Physics, Humboldt-Universit\"at zu Berlin, Newtonstr. 15, 12489 Berlin, Germany;}
\affiliation{Bernstein Center for Computational Neuroscience Berlin, Philippstr. 13, 10115 Berlin, Germany}
%\thispagestyle{empty}
%-----------------------------------------------------------------------

%-----------------------------------------------------------------------
\begin{abstract}
  We study networks of noisy phase oscillators whose nodes are characterized
  by a random degree counting the number of its connections. Both these degrees
  and the natural frequencies of the oscillators are distributed according to a
  given probability density. Replacing the randomly connected network by an
  all-to-all coupled network with weighted edges, allows us to formulate the
  dynamics of a single oscillator coupled to the mean field and to
  derive the corresponding Fokker-Planck equation. From the latter we
  calculate the critical coupling strength for the onset of
  synchronization as a function of the noise intensity, the frequency
  distribution and the first two moments of the degree distribution. 
  Our approach is applied to a dense small-world network
  model, for which we calculate the degree distribution. Numerical
  simulations prove the validity of the made replacement. We also test
  the applicability to more sparsely connected networks and formulate homogeneity
  and absence of correlations in the degree distribution as limiting
  factors of our approach.
\end{abstract}
\maketitle

\section{Introduction}
Synchronization phenomena have been observed in many different fields
such as the chaotic intensity fluctuations of coupled
lasers \cite{RTho94}, flashing swarms of fireflies \cite{BB76}, spiking dynamics of neurons
\cite{Dye91,EHRuSchimNe09,OlLiPoTo10,LucPo10}, pacemaker cells in the
heart \cite{DoDeTra89}, and the menstrual cycles of women living
together \cite{McC71,Wi92,YaSch06}, to mention only a few examples.
Without doubts, the phenomenon of synchronization is a central
mechanism in physics, chemistry, biology, and medicine as reported
impressively in various monographs \cite{PikRosKu03,Ta07,AniAstNeVaSchi07,BalJaPoSos10}.

The key aspect of our analysis is the phase description of the dynamics
\cite{LiGarNeiSchi04}. With regard to biological oscillators, the idea
of the phase description goes back to Winfree (1967) who motivated it
by the observation that the oscillators should be weakly coupled so
that no oscillator is ever perturbed far away from its limit cycle and
therefore amplitudes could be considered as fixed \cite{Win67}.

Kuramoto (1975, 1984) simplified the Winfree model by certain methods
of perturbation and averaging \cite{Kur75,Kur84}.  The Kuramoto model
has been shown to be a very successful approach to the problem of
synchronization \cite{OcG10}. By virtue of its simplicity the Kuramoto model
is analytically tractable. We take advantage of the simplicity to the
greatest extent, but at the same time we are determined to make the
model more realistic, in particular with regard to neural networks.
Therefore, we add two features: (i) we consider complex networks instead
of all-to-all connectivity and (ii) we study noise in the phase model. The
first feature takes account of the fact that many real-world networks,
such as neural systems, are complex networks ``par excellence''
\cite{SpChKaHil04}, while the second feature incorporates experimentally
observed stochastic processes, such as the random synaptic inputs from
other neurons \cite{LiGarNeiSchi04}.

For an ensemble of Kuramoto oscillators, both features
have been considered separately in the literature: complex networks in
\cite{HCK02,Ichi04,ResOttHu05} and noise in
\cite{Sak88,StrMir91,ZaNeFeSch03}. For an overview on nonlinear
dynamics in complex networks we recommend
\cite{AlBar02,New03,BocLaMoChHw06,ArDiazKuMoZh08} and for an overview
on the Kuramoto model we further recommend \cite{Str00,Ott02,AcBoViRiSp05}.

Referring to \cite{Ichi04}, we apply an approximation technique, in which 
we focus on the degrees of the nodes, i.e. the \colr{numbers} of connections
leading to and emanating from a node.
The degree distribution could be given ``a priori'' and then
defines the network or it is determined by counting the frequencies
of the edges for every node of the large network
under consideration.

As a result of the approximation, the complex
network is replaced by an all-to-all coupled one with weighted
edges, which mimics the original complex structure. This procedure
originates a mean-field like description that enables us to derive both the
effective Langevin equation for a single oscillator inside the random
network and the corresponding nonlinear Fokker-Planck equation (FPE).
Moreover, the latter gives us the possibility to calculate the critical coupling
strength that marks the onset of synchronization in the network.

The work is organized as follows. In section \ref{conception} we
present our extended Kuramoto model and in section \ref{approx_tech}
we explain the approximation technique. In section \ref{nfpe} we
derive the nonlinear FPE and in \ref{analyt_deriv} the critical
coupling strength. This result is compared with numerical simulations on
small-world networks (section \ref{simulations}),
\colr{and the applicability to other networks is discussed (section \ref{beyond}).}
In the last section \ref{conclusion} we give concluding remarks.

\section{Conception of Our Model}
\label{conception}
We refer to the original Kuramoto model \cite{Kur75,Kur84} in the following way:
\begin{equation}
  \dot{\phi}_i(t)=\omega_i+\frac{\varkappa}{N}\sum_{j=1}^{N}A_{ij}\sin\left(\phi_j(t)-\phi_i(t)\right)+\xi_i(t),\ i=1,\ldots,N\ ,
\label{ourmodel}
\end{equation}
\colr{with the number of oscillators $N$, the phase $\phi_i(t)$ at time $t$ and natural frequency
  $\omega_i$ of oscillator $i$.  The coupling strength is denoted by
  $\varkappa$ and $A_{ij}$ are the elements of the adjacency
  matrix. If the nodes $i$ and $j$ are not connected, then $A_{ij}=0$,
  otherwise $A_{ij}=1$. Self-coupling is excluded in weighted networks.
  The adjacency matrix further yields the individual degrees:
\begin{equation}
k_i=\sum_{j=1}^{N} A_{ij},\ i=1,\ldots,N.
\label{degrees}
\end{equation}
The natural frequencies and individual degrees are 
distributed according to a joint probability density $P(\omega,k)$.}

Remember that the indegree or the outdegree of node $i$ are defined as the 
\colr{number of edges pointing to} or emanating from node $i$, respectively
\cite{Wu07}. Since we consider undirected networks, the adjacency
matrix is symmetric and the indegrees are equal to the outdegrees.
Hence, we refer only to the degree $k_i$ of node $i$.

The division by $N$ in the coupling term of Eq. \eqref{ourmodel} is not
always the correct normalization for complex networks, because it might
not lead necessarily to an intensive coupling term. That is the case,
if the adjacency matrix does not scale with the system size. Then
one has to take the maximum degree instead \cite{ArDiazKuMoZh08}.

The functions $\xi_i(t),\ i=1,\ldots,N$, stand for sources of
independent white noise processes \colr{that act on the natural
frequencies as a stochastic force. They satisfy
\begin{equation}
\begin{aligned}
\langle\xi_i(t)\rangle&=0,\\
\langle\xi_i(t)\xi_j(t')\rangle&=2D\delta_{ij}\delta(t-t')\,.
\label{noise}
\end{aligned}
\end{equation}}
\colr{Hence, a single nonnegative parameter $D$ scales the noise. It is its intensity and we assume
that it is independent of the system size.} The angular brackets denote an average over different
realizations of the noise.

\section{Approximation Technique}
 \label{approx_tech}

The idea is to adopt a combinatorial point of view in order to mimic
the complex network by a weighted fully connected network with a new
adjacency matrix $\tilde{A}_{ij}$ which should resemble the original structure defined by the
given set of degrees $k_i,\ i=1,...,N$. We require that the
approximating weights $\tilde A_{ij}$ conserve the degrees of the
original network (cf. Eq. \eqref{degrees}), i.e.
\begin{equation}
  k_i\stackrel{!}=\sum_{j=1}^{N} \tilde A_{ij}\ \text{ and }\ k_j\stackrel{!}=\sum_{i=1}^{N} \tilde A_{ij},
\label{sumAij}
\end{equation}
where we assume again an undirected network. Eq. \eqref{sumAij} also means that the new matrix scales in the same way as the old one, if the system size is changed.

\begin{figure}
\centering
\includegraphics[width=0.9\linewidth]{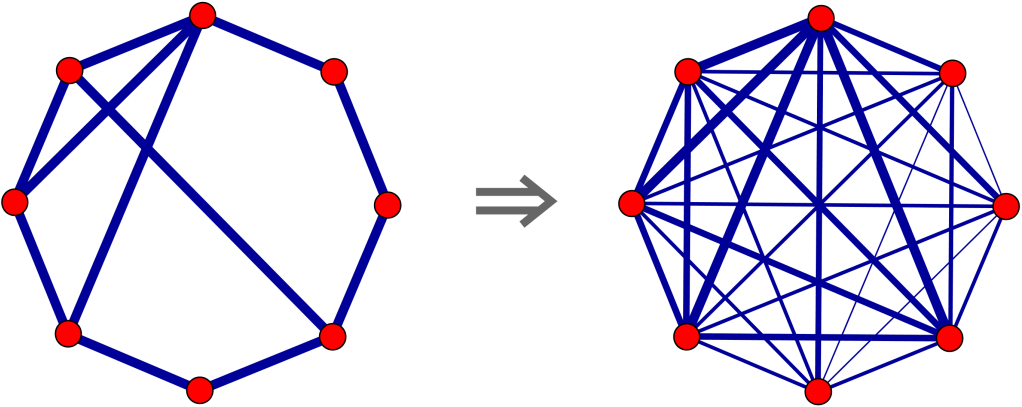}
\caption[The effect of our approximation on a ring network
of eight symmetrically coupled oscillators.]{(Color online) The effect of our
  approximation on a ring network of eight symmetrically
  coupled oscillators. On the left-hand side the original complex network
  is shown and on the right-hand side the approximate network is shown, 
where the thickness of the edges is chosen approximately proportional to the coupling strength.
For reasons of clarity, self-coupling is not visualized.}
\label{ourapprox}
\end{figure}

\colr{In order to construct the approximating weight between nodes $i$ and $j$,
we treat the original complex network as a random network and assume
that the coupling strength is proportional to the 
probability that a node with degree $k_i$ couples to a node with degree $k_j$.
If the degrees are uncorrelated \cite{Ichi04,BaBartVesp08}, then
$\tilde{A}_{ij}$ is equal to the degree $k_i$ times the degree $k_j$
normalized by the total number of all degrees in the network:}
\begin{equation}
  \tilde A_{ij}\,=\, k_i \, \frac{k_j}{\sum_{l=1}^{N} k_{l}}.
\label{newA}
\end{equation}
Obviously, it defines a symmetric matrix since the same expression is
found using the same arguments for the weight of the edge
connecting the $j$-th with the $i$-th node.
It is easy to see that the new matrix conserves the local degree
structure as required in Eq. \eqref{sumAij} and the scaling of the
adjacency matrix. We illustrate our approximation in Fig.
\ref{ourapprox}.

Inserting Eq. \eqref{newA} into Eq. \eqref{ourmodel} yields
\begin{equation}
  \dot{\phi}_i(t)=\omega_i+\frac{\varkappa }{N}\frac{k_i}{\sum_{l=1}^{N}k_l}\sum_{j=1}^{N}k_j\sin\left(\phi_j(t)-\phi_i(t)\right)+\xi_i(t),\ i=1,\ldots,N.
\label{ourmodelapprox}
\end{equation}
This is the approximative form of our model, which appears to be
analytically tractable. 

\colr{Let us make the following definition (cf. Ref. \cite{Ichi04}):}
\begin{equation}
  r(t) \mathrm e^{i\Theta(t)}:=\frac{\sum_{j=1}^N k_j \mathrm e^{i\phi_j(t)}}{\sum_{j=1}^N k_j}.
\label{orderp_ourmodel}
\end{equation}
\colr{The quantity $r(t)$ is an order parameter, because for a
population of $N\rightarrow\infty$ completely asynchronous oscillators, $r(t\rightarrow\infty)$ vanishes,
while the onset of synchronization is marked by $r(t\rightarrow\infty)>0$ (the completely phase
synchronized state corresponds to $r(t\rightarrow\infty)=1$).}

By multiplying Eq. \eqref{orderp_ourmodel} by $\mathrm e^{-i\phi_i(t)}$ and by considering only the imaginary parts, 
we can rewrite Eq. \eqref{ourmodelapprox} as an effective
one-oscillator description, where the common time-dependent phase
$\Theta(t)$ and amplitude $r(t)$ are averaged over all the nodes according to Eq. \eqref{orderp_ourmodel}:
\begin{equation}
  \dot{\phi}_i(t)=\omega_i+ r(t)\varkappa \frac{k_i}{N}\sin(\Theta(t)-\phi_i(t))+\xi_i(t),\ i=1,\ldots,N.
\label{ourmodelapproxrew}
\end{equation}
In other words, our approximation is equivalent to a mean-field
approximation with the mean field amplitude $r(t)$ and phase
$\Theta(t)$ defined in Eq. \eqref{orderp_ourmodel}. 
\colr{All oscillators are statistically identical and differ by 
$\omega_i$ and $k_i$ only. They are coupled to the
mean field with a characteristic strength that is proportional to
the individual degree $k_i$, which can be seen as a property of node $i$.
Such an assignment is similar to a characterization of subpopulations 
through coupling strengths \cite{HoStr11}.}

\section{Derivation of the Nonlinear Fokker-Planck Equation}
\label{nfpe}
The solution of the system of $N$ coupled Langevin equations in Eq.
\eqref{ourmodelapprox} is a Markov process. It can also be formulated by the help of the transition probability $\mathcal
P\left(\boldsymbol{\phi},t|\boldsymbol{\phi}^0,t^0;\boldsymbol{\omega},\boldsymbol{k}\right)$ that
describes the evolution of the phases from time $t^0$ to time $t>t^0$.
Therein, respectively the vector $\boldsymbol{\phi}=\left(\phi_1,\ldots,
  \phi_N\right)$ contains the phases of the $N$ oscillators at time $t$ and
$\boldsymbol{\phi}^0=\left(\phi_1^0,\ldots, \phi_N^0\right)$ at initial
time $t^0$. The vectors $\boldsymbol{\omega}=\left(\omega_1,\ldots,
  \omega_N\right)$ and $\boldsymbol{k}=\left(k_1,\ldots,
  k_N\right)$ contain the natural frequencies of the oscillators and the 
degrees of all the nodes in the network, respectively. The individual frequencies
$\boldsymbol{\omega}$ and degrees $\boldsymbol{k}$ are initially chosen
at time $t=t^0$ and they stay fixed during the whole evolution. 
We also assume that the initial phases $\phi_i^0$ are identically and independently distributed. 

The transition probability $\mathcal P\left(\boldsymbol{\phi},t|\boldsymbol{\phi}^0,t^0;\boldsymbol{\omega},\boldsymbol{k}\right)$
for $N$ oscillators satisfies a linear
Fokker-Planck equation (FPE)
\begin{equation}
  \frac{\partial\mathcal P}{\partial t}\, =\, - \sum_i\frac{\partial}{\partial\phi_i}\left(\omega_i \mathcal P + k_i \frac{\varkappa}{N}\sum_{j=1}^{N}\frac{k_j}{\sum_{l=1}^{N} k_l}\sin\left(\phi_j-\phi_i\right)\mathcal P\right)\,+\,D\sum_{i}\frac{\partial^2\mathcal P}{\partial\phi_i^2}
\label{FPE}
\end{equation}
and is subject to the initial condition
\begin{equation}
  \mathcal P\left(\boldsymbol{\phi},t^0|\boldsymbol{\phi}^0,t^0;\boldsymbol{\omega},\boldsymbol{k}\right)=\delta^N\left(\boldsymbol{\phi}-\boldsymbol{\phi}^0\right).
\end{equation}
Conditioned and joint probabilities satisfy
\begin{equation}
   \mathcal P\left(\boldsymbol{\phi},t|\boldsymbol{\phi}^0,t^0;\boldsymbol{\omega},\boldsymbol{k}\right)\mathcal P\left(\boldsymbol{\phi}^0,t^0;\boldsymbol{\omega},\boldsymbol{k}\right)=\mathcal P\left(\boldsymbol{\phi},t;\boldsymbol{\omega},\boldsymbol{k}|\boldsymbol{\phi}^0,t^0\right)\mathcal P\left(\boldsymbol{\phi}^0,t^0\right)\, .
 \label{bayes}
\end{equation}
As already stated in section \ref{conception}, the natural frequencies and degrees are
distributed according to the joint probability density $P(\omega,k)$. 
The initial phases of the oscillators are further supposed to be chosen
independently of $\boldsymbol{\omega}$ and $\boldsymbol{k}$. 
Therefore, we have the joint distribution of initial phases,
edges and frequencies at initial time $t^0$ factorizing as
\begin{equation}
  \mathcal P\left(\boldsymbol{\phi}^0,t^0;\boldsymbol{\omega},\boldsymbol{k}\right)\,=\, \prod_{i=1}^N \,P\left(\omega_i,k_i\right) \, \mathcal P\left(\phi_i^0,t^0\right)\,.
\label{initial}
\end{equation}
The joint distribution at a later time $t$ becomes independent of the
initial distribution by integrating Eq. \eqref{bayes} over the initial phases:
\begin{equation} {\mathcal
    P}\left(\boldsymbol{\phi},t;\boldsymbol{\omega},\boldsymbol{k}\right)=\int\mathrm d{\phi_1^0}\ldots\mathrm d{\phi_N^0}\ \mathcal
    P\left(\boldsymbol{\phi},t|\boldsymbol{\phi}^0,t^0;\boldsymbol{\omega},\boldsymbol{k}\right)\prod_{i=1}^N
    P\left(\omega_i,k_i\right)\mathcal
    P\left({\phi_i^0},t^0\right)\,.
\label{P}
\end{equation}

Again, we aim at reducing the description to an effective one-oscillator
picture. As does the transition
probability, the distribution Eq. \eqref{P} obeys the FPE \eqref{FPE}.
A reduced joint distribution $\rho_n$ with $n<N$ for $n$ phases with
frequencies $\omega_1 \ldots \omega_n$ and degrees $k_1,\ldots
k_n$ is defined in the usual way \cite{CraDa99}:
\begin{equation}
\begin{aligned}
  \rho_n & \left(\phi_1,\ldots, \phi_n,t;\omega_1,\ldots, \omega_n,k_1,\ldots, k_n\right)\\
  &:=\int\mathrm d\phi_{n+1}\ldots\mathrm d\phi_N\int\mathrm
  d\omega_{n+1}\ldots\mathrm d\omega_N\int\mathrm d
  k_{n+1}\ldots\mathrm d k_N\ \mathcal
  P\left(\boldsymbol{\phi},t;\boldsymbol{\omega},\boldsymbol{k}\right).
\end{aligned}
\end{equation}
For the important case of a single oscillator $n=1$ imbedded in the
network, we integrate Eq. \eqref{FPE} over the remaining $N-1$ phases
$\left(\phi_2,\ldots, \phi_N\right)$, frequencies
$\left(\omega_2,\ldots, \omega_N\right)$ and degrees
$\left(k_2,\ldots, k_N\right)$:
\begin{equation}
\begin{aligned}
  &\frac{\partial\rho_1}{\partial t}=-\frac{\partial}{\partial \phi_1}\omega_1\rho_1 + D\frac{\partial^2\rho_1}{\partial\phi_1^2}\,- \\
  &-\frac{\partial}{\partial
    \phi_1}\left[\frac{\varkappa}{N}\frac{k_1\left(N-1\right)}{\sum_{l}k_l}\int\mathrm
    d \phi_2\int\mathrm d \omega_2\int\mathrm d k_2
    \sin\left(\phi_2-\phi_1\right)k_2\rho_2\left(\phi_1,\phi_2,t;\omega_1,\omega_2,k_1,k_2\right)\right]\ ,
\end{aligned}
\end{equation}
where we use identity of all the oscillators in the statistical sense. 
As a result the binary interaction of the phase
oscillators looks in the usual way and relates hierarchically to
reduced probabilities with a larger number $n>2$. However, the interaction
appears as averaged over the frequency-degree distribution (cf. Eq. \eqref{order_fpe}).

In the thermodynamic limit $N\rightarrow\infty$ the ratio $\sum_{l}k_l/(N-1)$
becomes the average degree. \colr{Moreover, the correlation of phases between any two
oscillators can be discarded \cite{CraDa99} within the scope of the mean-field approximation,
Eq. \eqref{ourmodelapprox}:
\begin{equation}
  \rho_2\left(\phi_1,\phi_2,t;\omega_1,\omega_2,k_1,k_2\right)=\rho_1\left(\phi_1,t;\omega_1,k_1\right)\rho_1\left(\phi_2,t;\omega_2,k_2\right)\,.
\end{equation}
Such a decoupling has also been used in \cite{Ichi04,ResOttHu05,Lee05,BaBartVesp08}
and it is an omnipresent approximation in the theory of phase transitions \cite{Sta71}.
A completely rigorous discussion about the general conditions has been proposed in \cite{AcBoViRiSp05}
by the help of path integral methods. We point out
that this effective decoupling is fully equivalent to the former
description by the help of the stochastic differential equation for the
single phase oscillator imbedded in the mean field (cf. Eq. \eqref{ourmodelapproxrew})}.

Let us return to the conditional probability density
\begin{equation}
\rho_1\left(\phi_1,t|\omega_1,k_1\right)=\frac{\rho_1\left(\phi_1,t;\omega_1,k_1\right)}{P\left(\omega_1,k_1\right)}\,,
\label{cond_1}
\end{equation}
by which we get eventually a nonlinear Fokker-Planck equation for
the one-oscillator probability density $\rho_1$:
\begin{equation}
\begin{aligned}
  \frac{\partial \rho_1\left(\phi_1,t|\omega_1,k_1\right)}{\partial
    t}\,=\,-\,\frac{\partial}{\partial\phi_1}\left[  v(\phi_1,t) \, \rho_1\left(\phi_1,t|\omega_1,k_1\right)\right]\,+\,D\frac{\partial^2
    \rho_1\left(\phi_1,t|\omega_1,k_1\right)}{\partial\phi_1^2}\ .
\label{fpe}
\end{aligned}
\end{equation}
Therein the mean increment of the phase per unit time reads
\begin{equation}
\begin{aligned}
  v(\phi_1,t)\,=\,\omega_1+r\tilde\varkappa k_1\sin(\Theta-\phi_1)\,.
  \label{v}
\end{aligned}
\end{equation}
It depends on the density $\rho_1(\phi_1,t|\omega_1,k_1)$ via the order parameter
\begin{equation}
\begin{aligned}
  r\mathrm e^{i\Theta}&=\frac{1}{\langle
    k\rangle}\int_{0}^{2\pi}\mathrm d\phi_2\int_{-\infty}^{+\infty}\mathrm d\omega_2\int_{m}^{\infty}\mathrm dk_2\ \mathrm
  e^{i\phi_2}\ \rho_1\left(\phi_2,t|\omega_2,k_2\right)\ k_2\ 
  P\left(\omega_2,k_2\right)\, .
  \label{order_fpe}
\end{aligned}
\end{equation}
Here $\tilde\varkappa:=\varkappa/N$ and $m\geq1$ is the lowest
possible degree in our network. \colr{Eqs. \eqref{fpe}-\eqref{order_fpe}
constitute again an effective one-oscillator description as
Eqs. \eqref{orderp_ourmodel}, \eqref{ourmodelapproxrew}.} Both
descriptions are equivalent for all possible pairs $\omega_i$ and
$k_i$\colr{, the difference is that in this section the equations
are derived for the thermodynamic limit}.
We remark that our expressions are reminiscent of the
equations in the work of Ichinomiya \cite{Ichi04}, who considers
uncorrelated random networks without noise. \colr{We emphasize that the
above averaging requires a coarse grained statistical approach to all the
nodes. Since each of the $N$ oscillators is coupled to the
mean field with a characteristic strength,
we refer to the ``weighted mean field'', which makes sense if the
initial conditions of all oscillators are identically distributed.}
We also note that the conditional
probability density
$\rho_1\left(\phi_1,t|\omega_1,k_1\right)$ in Eq. \eqref{cond_1} is
averaged over the initial conditions. If one would formulate the
theory for the transition probability
$\rho_1\left(\phi_1,t|\phi^0,t^0,\omega_1,k_1\right)$ the nonlinear
FPE \eqref{fpe} would be the same.  A difference would occur in the
definition of the order parameter, where one has to remember that the
definition \eqref{order_fpe} includes the average over the initial
conditions as it was performed in Eq. \eqref{initial}.

\section{The Critical Coupling Strength}
\label{analyt_deriv}
To underline the statistical identity of the oscillators in the
one-oscillator description, we omit the indices and proceed with
variables $\phi, \omega, k$. Taking into account the normalization
condition
\begin{equation}
\int_0^{2\pi}\rho(\phi,t|\omega,k)\mathrm d\phi=1\,, \,~~~ \forall\ \omega,k,t\ ,
\label{rho_norm}
\end{equation}
the completely asynchronous stationary state is given by
\begin{equation}
  \rho_0(\phi | \omega,k)=\frac{1}{2\pi}\,=\text{const.}\,, \,~~~ \forall\ \omega,k,t
\end{equation}
with order parameter $r=r_0=0$ and undefined phase. It is easy to
see that $\rho_0$ is at least one of the possible asymptotic solutions of
the nonlinear FPE \eqref{order_fpe}.

In the following, we will perform a linear stability analysis of this
incoherent solution. In doing so, we refer to the work of Strogatz and
Mirollo \cite{StrMir91}, who \colr{also study a nonlinear FPE in
order to} investigate the linear stability of the
incoherent state for an unweighted all-to-all connectivity. The
critical condition, where the incoherent solution loses its stability,
is equivalent to the onset of synchronization we are searching for.
To start with, we consider the evolution of a small perturbation of
the incoherent state:
\begin{equation}
\rho(\phi,t|\omega,k)=\frac{1}{2\pi}+\epsilon\delta\rho(\phi,t|\omega,k),\ \epsilon\ll 1.
\label{linear_perturb}
\end{equation}
The normalization condition \eqref{rho_norm} implies
\begin{equation}
  \epsilon\int_0^{2\pi}\delta\rho(\phi,t|\omega,k)\mathrm d\phi=0\ \forall\ \omega,k,t\ ,
\label{eta_2piper}
\end{equation}
so $\delta\rho(\phi,t|\omega,k)$ is $2\pi$-periodic in $\phi$. Now we substitute Eq. \eqref{linear_perturb} into Eq. \eqref{fpe},
\begin{equation}
  \epsilon\frac{\partial\delta\rho}{\partial t}=-\frac{\partial}{\partial\phi}\left[\left(\frac{1}{2\pi}+\epsilon\delta\rho\right)v\right]+\epsilon D\frac{\partial^2\delta\rho}{\partial\phi^2}\ .
\end{equation}
The amplitude becomes $r(t)=\epsilon \delta r(t)$, where $\delta r(t)$ is given by substituting $\rho$ by $\delta\rho$ in Eq. \eqref{order_fpe}.
The FPE \eqref{fpe} linearized in the lowest order in $\epsilon$ then reads
\begin{equation}
  \frac{\partial\delta\rho}{\partial t}=-\omega\frac{\partial\delta\rho}{\partial\phi}+\frac{\tilde\varkappa k}{2\pi}\delta r\cos(\Theta-\phi)+D\frac{\partial^2\delta\rho}{\partial\phi^2}\ .
\label{o_eps}
\end{equation}
Since $\delta\rho$ is a real number and $2\pi$-periodic in $\phi$ (cf. Eq. \eqref{eta_2piper}), it may be expanded as
\begin{equation}
\delta\rho(\phi,t|\omega,k)=\frac{1}{2\pi}\sum_{l=1}^{\infty}\left\{c_l(t|\omega,k)\mathrm e^{il\phi}+c_l^{\ast}(t|\omega,k)\mathrm e^{-il\phi}\right\}\ .
\label{expansion}
\end{equation}
Due to Eq. \eqref{eta_2piper}, $c_0(t|\omega,k)$ vanishes. The first nonvanishing coefficients, which constitute the so-called 
fundamental mode, determine the deviations of the mean field in first order of small $\epsilon$:
\begin{equation}
  \delta r\mathrm e^{i\Theta}=\frac{1}{\langle k\rangle}\int_{-\infty}^{+\infty}\mathrm d\omega'\int_{m}^{\infty}\mathrm dk' c_1^{\ast}(t|\omega',k')k'P\left(\omega',k'\right)\ .
\label{meanf_and_fundm}
\end{equation}
In order that the incoherent state becomes unstable and a synchronization
process starts, $c_1(t|\omega,k)$ has to grow, so that as
a consequence the order parameter $r(t)$ grows as well.

Multiplication of the last equation by $\mathrm e^{-i\phi}$ and considering only
the real parts, yields
\begin{equation}
\delta r\cos(\Theta-\phi)=\frac{1}{2\langle k\rangle}\left(\int_{-\infty}^{+\infty}\mathrm d\omega'\int_{m}^{\infty}\mathrm dk' c_1(t|\omega',k')k'P\left(\omega',k'\right)\right)\mathrm e^{i\phi}+\operatorname{c.c.}\ .
\label{rew_exp}
\end{equation}
Now we insert Eqs. \eqref{expansion}-\eqref{rew_exp} into Eq.
\eqref{o_eps} and consider only the coefficients with $\mathrm
e^{i\phi}$. Afterwards, one finds the evolution equation for the
fundamental mode $c_1(t|\omega,k)$:
\begin{equation}
\frac{\partial c_1}{\partial t}=-(D+i\omega)c_1+\frac{\tilde\varkappa k}{2\langle k\rangle}\int_{-\infty}^{+\infty}\mathrm d\omega'\int_{m}^{\infty}\mathrm dk' c_1(t|\omega',k')k'P\left(\omega',k'\right)\ .
\label{evolution_fmode}
\end{equation}
Since this is a partial differential equation without mixed derivatives, we make a separation ansatz:
\begin{equation}
c_1(t|\omega,k)\equiv b(\omega,k)\ \mathrm e^{\lambda t},\ \lambda\in\mathbb C.
\label{ansatz_c1}
\end{equation}
So we obtain
\begin{equation}
\lambda b=-(D+i\omega)b+\frac{\tilde\varkappa k}{2\langle k\rangle}\int_{-\infty}^{+\infty}\mathrm d\omega'\int_{m}^{\infty}\mathrm dk' b(\omega',k')k'P\left(\omega',k'\right)\ .
\label{eq_without_f}
\end{equation}
The right-hand side is a linear transformation and it can be shown that we only have to calculate its point spectrum in order to obtain the critical coupling strength \cite{StrMir91}.

The second term on the right-hand side of Eq. \eqref{eq_without_f} equals the degree $k$ times a constant which we call $B$ in the following, so that $b(\omega,k)$ is given by
\begin{equation}
b(\omega,k)=\frac{B\cdot k}{\lambda+D+i\omega}\ .
\end{equation}
Hence, we can set up the following self-consistent equation:
\begin{equation}
B=\frac{\tilde\varkappa}{2\langle k\rangle}\int_{-\infty}^{+\infty}\mathrm d\omega'\int_{m}^{\infty}\mathrm dk'\frac{B\cdot k'^2}{\lambda+D+i\omega'}P\left(\omega',k'\right)\ .
\label{selfcons}
\end{equation}
This relation holds, if all the oscillators are identical in the
statistical sense, see section \ref{nfpe}. \colr{We remark that
  Restrepo, Ott and Hunt (2005) derived a similar formula where
  fluctuations emerge due to finite-size effects \cite{ResOttHu05}. 
  We have defined in Eq. \eqref{noise} the value $D$. As outlined,
  it does not vanish in the thermodynamic
  limit. In Eq. \eqref{selfcons} the first integral exists for all
  $\operatorname{Re}(\lambda)>-D$ \cite{Ott02}}. The solution $B=0$ is not allowed,
because otherwise $b(\omega,k)$ and as a consequence $c_1(t|\omega,k)$
would vanish. $c_1(t|\omega,k)=0$ however is not considered as an
eigenfunction (it is a trivial solution). So $B$ is canceled out and
we get
\begin{equation}
1=\frac{\tilde\varkappa}{2\langle k\rangle}\int_{-\infty}^{+\infty}\mathrm d\omega'\int_{m}^{\infty}
\mathrm dk'\frac{k'^2}{\lambda+D+i\omega'}P\left(\omega',k'\right)\ .
\label{eveq}
\end{equation}
On the analogy of the proof in \cite{MiStr90}, one can show that there exists only one solution for $\lambda$
and that it has to be a real number. The proof makes use of the assumption \colr{that, with regard to the
$\omega$-dependency, $P(\omega,k)$ has a single maximum at frequency $\omega=0$ (due to the rotational symmetry
in the model the maximum can be located like this) and is symmetric.}

In summary the eigenvalue $\lambda$ is given by
\begin{equation}
1=\frac{\tilde\varkappa}{2\langle k\rangle}\int_{-\infty}^{+\infty}\mathrm d\omega'\int_{m}^{\infty}\mathrm dk'\frac{(\lambda+D)\cdot k'^2}{(\lambda+D)^2+\omega'^2}P\left(\omega',k'\right)\ .
\label{eigenvalue}
\end{equation}
Note that this equation can only be fulfilled if $\lambda>-D$, because otherwise the right-hand side would be nonpositive. Thus, the eigenvalue $\lambda$ is \colr{strictly positive} and the incoherent solution cannot be linearly stable, if the noise intensity vanishes.

At the critical condition $\lambda=\lambda_c=0$ the incoherent solution loses its stability: if $\lambda>0$ the fundamental
mode $c_1$ of the perturbation $\delta\rho$ of the incoherent solution is linearly unstable. It grows exponentially with
time $\sim\mathrm e^{\lambda t}$ and so does the order parameter $r(t)$ (cf. Eq. \eqref{meanf_and_fundm}). Hence, the critical
coupling strength $\varkappa_c$ for the onset of synchronization reads
\begin{equation}
  \varkappa_c=2 N \langle k\rangle\left[\int_{-\infty}^{+\infty}\mathrm d\omega'\int_{m}^{+\infty}\mathrm dk'\frac{D\cdot k'^2}{D^2+\omega'^2}P(\omega',k')\right]^{-1}\ .
\end{equation}
We emphasize that this equation is not valid in the noise-free case, where one has to take
the limit $\lambda\to 0^{+}$ in Eq. \eqref{eigenvalue} with $D=0$, by which previous results can be reproduced \cite{Ichi04,ResOttHu05} (compare Tab. \ref{fdiv_limiting}). 

In \cite{SonnSchiSa} we investigate the effects 
of correlations between the degrees $\boldsymbol{k}$ 
and the frequencies $\boldsymbol{\omega}$.
Here we continue with two separated distributions $g(\omega)$ and $P(k)$.
It is interesting to see that the critical coupling strength $\varkappa_c$ can then be written as a product of two functionals. The first one maps the degree distribution $P(k)$ to a real number via the first two moments,
\begin{equation}
f_{\rm{top}}[P]:= N \frac{\langle k\rangle}{\langle k^2\rangle}.
\label{top_func}
\end{equation}
We call this one the ``topology functional''. The second one maps the frequency distribution $g(\omega)$ to a real number via an integral that depends on the noise intensity $D$,
\begin{equation}
  f_{\rm{div}}(D)[g]:=2\left[\int_{-\infty}^{+\infty}\frac{D}{D^2+\omega'^2}g(\omega')\mathrm d\omega'\right]^{-1}.
  \label{func_div}
\end{equation}
We call this one the ``diversity functional''. In short:
\begin{equation}
\varkappa_c=f_{\rm{top}}[P]\cdot f_{\rm{div}}(D)[g].
\label{kappac}
\end{equation}

\section{Application to a Dense Small-World Network Model}
\label{simulations}
In this section we test the analytical expression for the critical
coupling strength as a function of diversity, noise and network topology for local
ring-like networks, where random shortcuts to other nodes
will be established. For this purpose, we compare the analytical result
with numerical simulations of networks with different size $N$.

\subsection{Derivation of the Topology Function}
First, we specify our dense small-world network model.
According to the definition in \cite{Pr98}, in dense networks of $N$ nodes
the total number of edges scales with $N^2$. Our dense small-world
networks are constructed as follows. Each oscillator is coupled to its
$K$ next neighbors in both directions of the ring network and $K$ is
given by
\begin{equation}
K=\left\lfloor\frac{\alpha}{2}(N-1)\right\rfloor,\ 0\leq\alpha\leq 1.
\label{Kalpha}
\end{equation}
Here the intensive variable $\alpha$ gives the fraction of all nodes
\colr{that are coupled through next neighbor connections}. $\alpha=0$ stands for
uncoupled nodes, whereas in case of $\alpha=1$ the network is fully
connected. We underline that the variance of $\alpha$ 
vanishes. The floor function $\lfloor .\rfloor$ gives the greatest
integer less than or equal to its argument\colr{, by which the continuous variable
$\alpha$ is mapped to the discrete variable $K$}. So far we obtain a regular network 
and due to self-coupling, the fixed degree in such regular networks is
$2K+1$. The corresponding degree distribution is equal
to the Kronecker symbol
\begin{equation}
  \label{eq:local_net}
  P_{\rm local}(k)\,=\,\delta_{k, 2K+1}\ .
\end{equation}
In our final simulations we choose $\alpha=0.05$. Hence, $2.5$
percent of possible edges are local regular connections.

\begin{figure}
\centering
\includegraphics[width=0.9\linewidth]{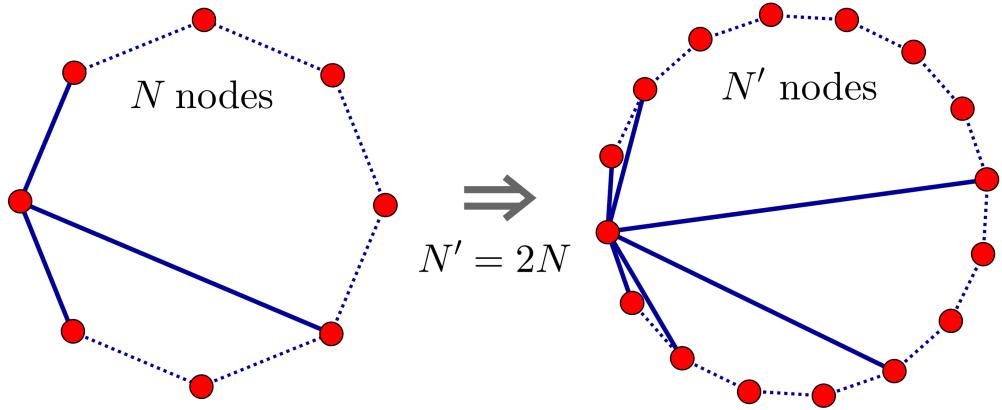}
\caption[Visualization of a dense small-world network
model.]{(Color online) Visualization of a dense small-world network model. If the
  system size $N$ is duplicated, the average number of connections is
  duplicated as well. The figure shows one marked node in the
  network.}
\label{dense}
\end{figure}

Besides these $(2K+1)\cdot N$ next neighbor couplings we introduce
shortcuts: each oscillator is coupled to the remaining other
$N-2K-1$ ones with a certain probability $p$ (see Fig. \ref{dense}
for illustration).
\colr{If $K$ equals zero the model generates Erd\H os-R\'enyi random networks.
The case $p=1$ leads to all-to-all connectivity other than in popular
sparse small-world network models \cite{WaStr98,NewWa99}.} The procedure
of adding shortcuts can be seen as a Bernoulli experiment with
probability $p$ of success \cite{AlBar02,New03,BocLaMoChHw06}, so that
the degree distribution $P(k)$ is given by the following binomial distribution:
\begin{equation}
P(k)\,=\, \binom{N-2K-1}{k-2K-1}p^{k-2K-1}\left[1-p\right]^{N-k}, ~~~~~ k>2K\,.
\label{degreedistr}
\end{equation}
\begin{figure}
\centering
\includegraphics[width=0.9\linewidth]{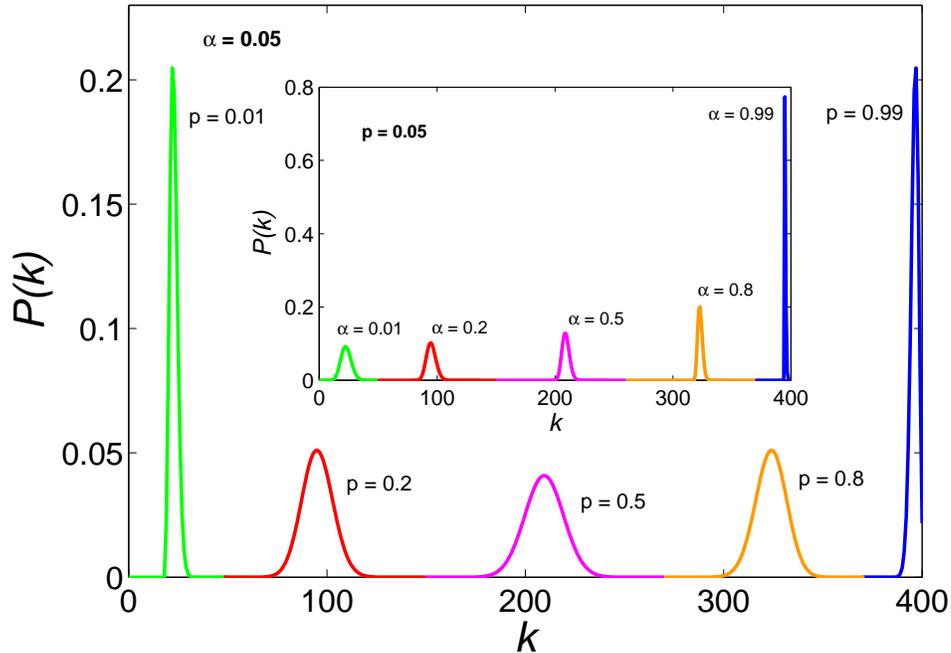}
\caption[]{\colr{(Color online) Degree distribution $P(k)$ of dense small-world networks
consisting of $400$ nodes depicted for different shortcut probabilities $p$
and fractions of regular connections $\alpha$ (inset). For the sake of clarity solid lines are chosen.}}
\label{Pk}
\end{figure}
Note that the minimum degree $k=2K+1$ corresponds to Eq.
\eqref{eq:local_net}. The first moment equals
\begin{equation}
  \langle k\rangle=2K+1+(N-2K-1)p  \approx (N-1)(\alpha+p-\alpha p)+1\ . 
\label{k1}
\end{equation}
In the last step we have used Eq. \eqref{Kalpha} by neglecting
the floor function as an approximation. For large $N$ the latter
expression approaches $\langle k \rangle \approx N(\alpha+p-\alpha p)$.

\begin{figure}
\begin{centering}
  \subfigure[$\ \alpha=0.052$]{\label{ftop_ndep_p}\includegraphics[width=0.49\linewidth]{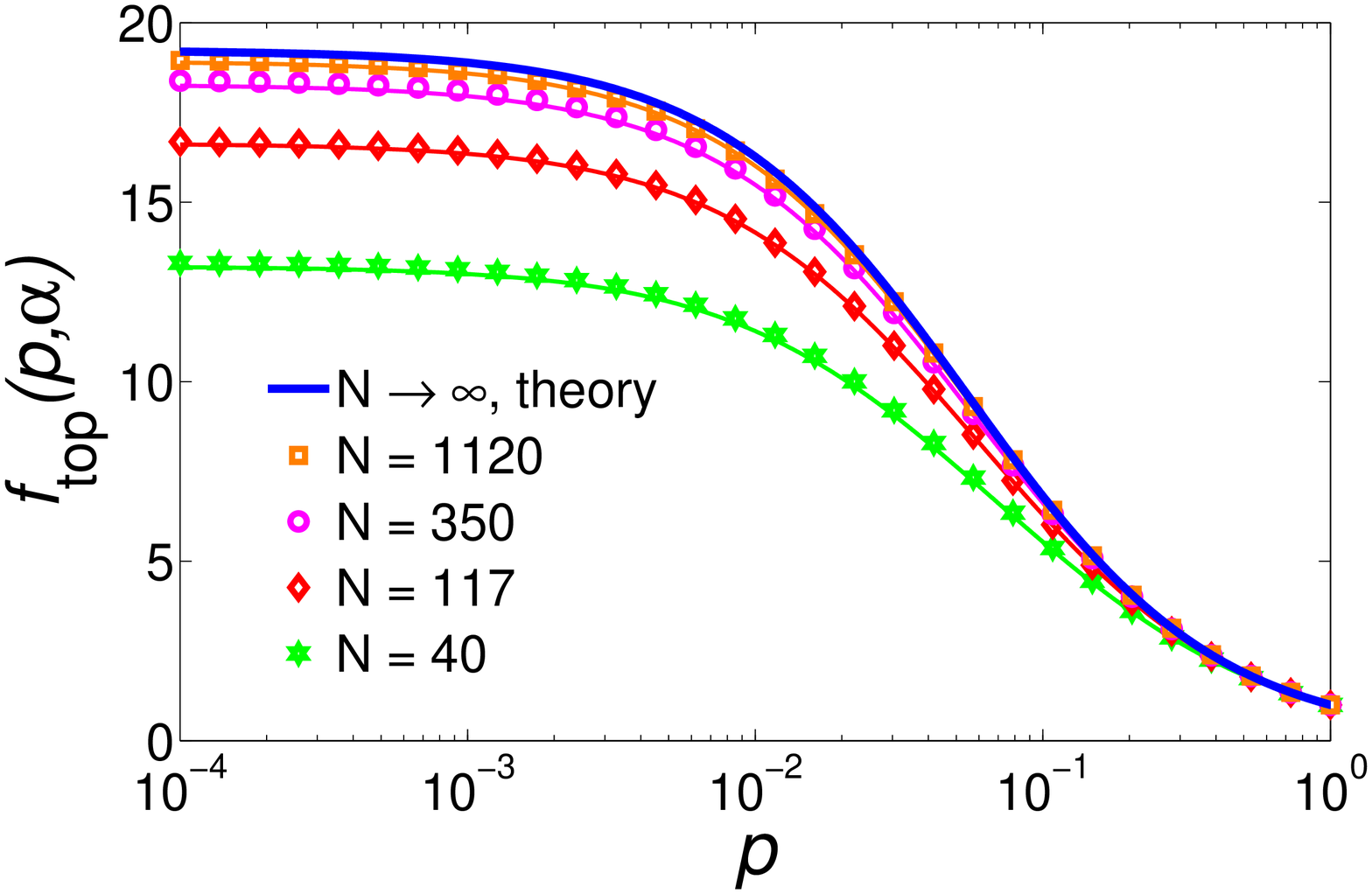}}
  \subfigure[$\ p=0.052$]{\label{ftop_ndep_a}\includegraphics[width=0.49\linewidth]{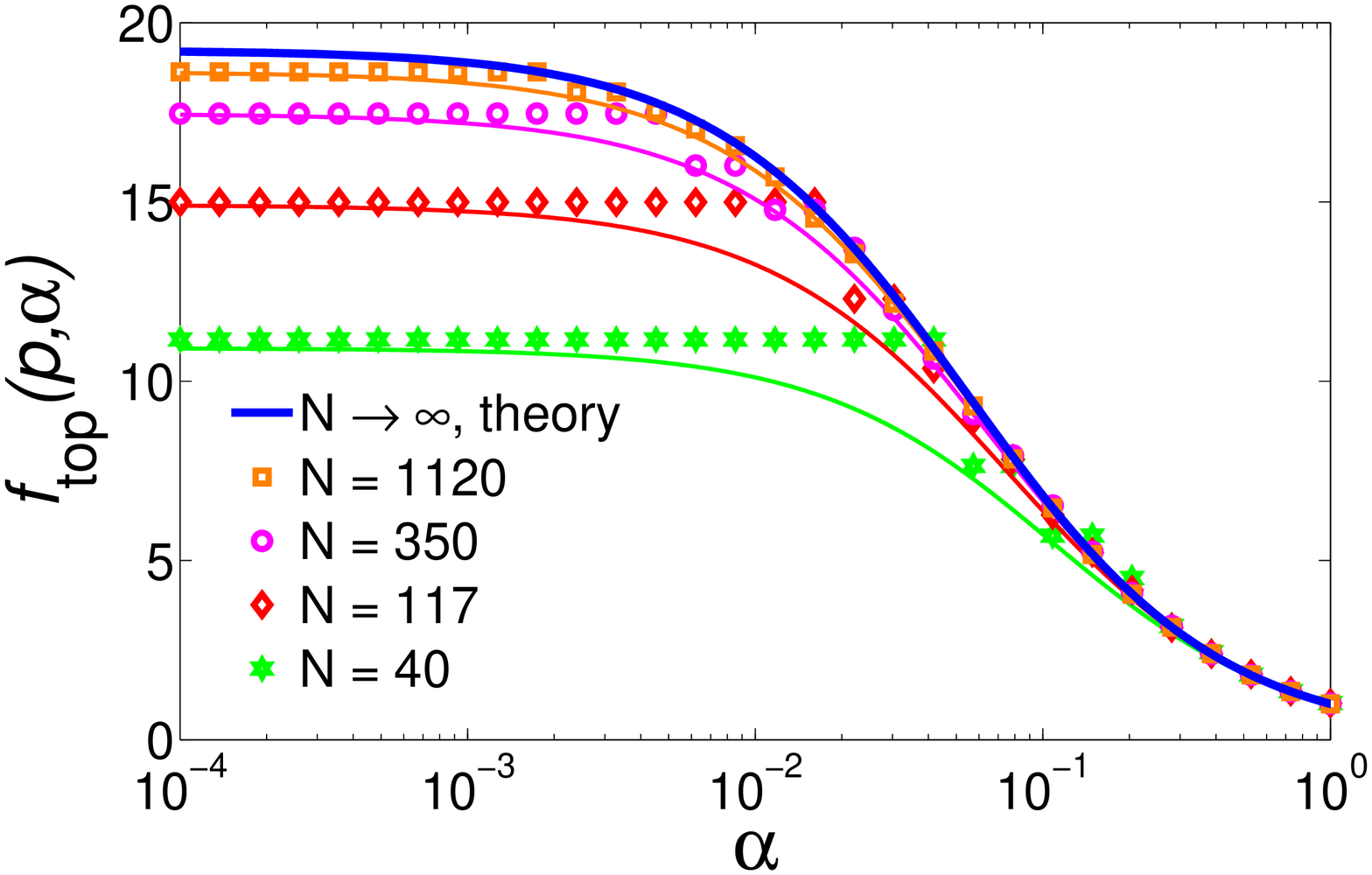}}
  \caption[The dependency of the topology function on the system size
  $N$ is shown.]{(Color online) Dependency of the topology function on
    the system size $N$: Solid lines are given by the theory. The
    thick line corresponds to the thermodynamic limit and slight lines
    are calculated according to Eqs. \eqref{k1} and \eqref{k2} for
    smaller systems. Markers show results of numerical calculations
    for indicated system sizes.}
  \label{ftop_ndep}
\end{centering}
\end{figure}

By simple substitutions one finds the second moment
\begin{eqnarray}
\langle k^2\rangle= \langle k \rangle^2 + (\langle k \rangle-2K-1)(1-p)\ ,
\label{k2}
\end{eqnarray}
where the second item equals the variance of $k$ in the binomial
distribution due to the randomness of the number of shortcuts. It
scales only linearly with the system size, whereas the first item grows with
$N^2$. Thus, for large systems the difference between the nonrandom
number of local connections and the number of shortcuts disappears, because the variability of the shortcuts does not count for large $N$. The second moment approaches $\langle k \rangle^2$ and it becomes symmetric in $\alpha$ and $p$. Hence, the topology function (cf. Eq. \eqref{top_func}) for
large $N$ results in
\begin{equation}
\left.f_{\rm{top}}(p,\alpha)\right|_{N\gg1}\approx\frac{N}{\langle k\rangle}\approx\frac{1}{\alpha+p-\alpha p}\ .
\label{approxftop}
\end{equation}
We obtain as limiting cases (compare Fig. \ref{ftop_ndep})
\begin{equation}
\left.f_{\rm{top}}(p,\alpha)\right|_{N\gg1}\approx\begin{cases}
 \ \ \ \ \ \ \frac{1}{\alpha}+\frac{\alpha-1}{\alpha^2}p+O(p^2),  & p\ll1, \\
 2-\alpha+(\alpha-1)p+O((1-p)^2),  & p\lesssim 1.
\end{cases}
\label{ftop_limiting}
\end{equation}
Due to the symmetry, we obtain qualitatively the same for $ \alpha\ll1$ and $\alpha\lesssim 1$, but with $p$ and $\alpha$ being interchanged. As expected, the topology function tends to unity for $p\rightarrow 1$
or $\alpha\rightarrow 1$, because in both cases the network becomes fully connected.

In Fig. \ref{ftop_ndep} the dependence of the topology function on $\alpha$ and $p$ is depicted.
Discrepancies between numerical calculations and theory in the right panel are due to the fact that
the number of coupled next neighbors is, of course, an integer. Rounding off in Eq. \eqref{Kalpha} 
leads to noticeable steps in $f_{\rm{top}}(p,\alpha)$ for smaller $N$.

We emphasize that by the help of Eqs. \eqref{k1} and \eqref{k2}, one
can find the exact expression for the topology function for arbitrary
system sizes.  Since it is a rather lengthy expression, we skip it in the 
text but we will use it in Fig. \ref{ftop_ndep}.
Moreover, as can be seen in Fig. \ref{ftop_ndep}, a system size 
$N$ of the order $O(100)$ is sufficient to obtain a dynamical
behavior that is comparable with the thermodynamic limit. Further, the critical
coupling strength $\varkappa_c$ has been derived in the thermodynamic
limit (see Eq. \eqref{kappac}). For this reason it is only
consistent to calculate the topology function in the limit
$N\rightarrow\infty$ as well.

\subsection{Comparison of the Critical Coupling Strength; Simulations vs. Theory}
\label{comparison}
In our simulations, the stochastic differential equations \eqref{ourmodel}
are integrated up to $t=600$ with time 
step $h=0.05$ by using the Heun scheme \cite{Man00}. The periods 
of the oscillators are $T\sim O(10)$, so our integrations
cover $O(10)$ periods. Moreover, in order to calculate statistical
equilibria, we discard the data up to $t=200$, by which transient effects are safely
avoided. The statistical equilibria are further
calculated as averages over at least $100$ different network
realizations. We emphasize that all the different network
configurations do not differ only in the configuration of the
connections, but the oscillators on the network differ as well: all
the natural frequencies and the initial values of the phases change
from one configuration to another one.

Hong, Choi and Kim (2002) consider the Kuramoto model without noise
on a small-world network, which is constructed according to the
Watts-Strogatz model \cite{WaStr98}, where shortcuts are the result of
rewiring the edges of the initial regular network with a certain
probability $p$ \cite{HCK02}. They calculate the order parameter
averaged over time $\langle\cdots\rangle$ and network realizations
$\left[\cdots\right]$ and use the following scaling form:
\begin{equation}
  s:=\left[\left\langle\left|\frac{1}{N}\sum_{j=1}^{N}\mathrm e^{i\phi_j}\right|\right\rangle\right]=N^{-\frac{\beta}{\nu}}F\left[(\varkappa-\varkappa_c)N^{\frac{1}{\nu}}\right]\ ,
\label{sync_par}
\end{equation}
where $F[.]$ is some scaling function. In particular it is found that $\beta$ and $\nu$ have the same values as in the
globally connected network, namely $\beta\approx\frac{1}{2}$ and $\nu\approx2$.

\begin{figure}
\begin{psfrags}
\psfragscanon
\psfrag{kappa}[t][t]{\color[rgb]{0,0,0}\setlength{\tabcolsep}{0pt}\begin{tabular}{c}$\varkappa$\end{tabular}}
\psfrag{kc }[t][t]{\color[rgb]{0,0,0}\setlength{\tabcolsep}{0pt}\begin{tabular}{c}$\varkappa_c$\end{tabular}}
\begin{center}
\includegraphics[width=0.49\linewidth]{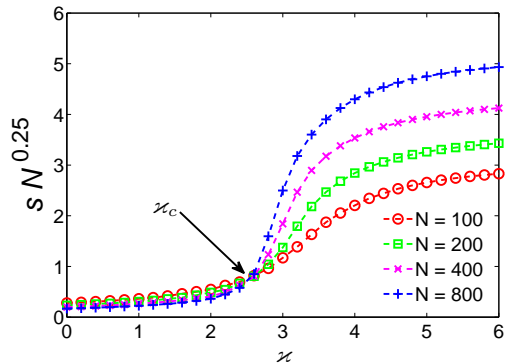}
\end{center}
\end{psfrags}
\caption[Determination of critical coupling strength]{(Color online)
  Numerical determination of critical coupling strength by plotting
  the order parameter $s$ times $N^{0.25}$ for
  different system sizes in dependence on the coupling strength.
  According to Eq. \eqref{sync_par} the intersection of the curves
  indicates the critical coupling strength. Parameter values here:
  $\alpha=0.05$, $p=0.1$, $D=0.05$, $\sigma=0.02$.}
\label{kappac_scal}
\end{figure}

The findings propose a finite-size scaling analysis to calculate the critical coupling strength, because at
$\varkappa=\varkappa_c$ the function $F[.]$ is independent of the system size $N$. By plotting
$s\cdot N^{0.25}$ as a function of $\varkappa$ for
various network sizes $N$, we can measure the critical coupling
strength $\varkappa_c$ as a well-defined intersection point, see Fig. \ref{kappac_scal}.

In order to compare the theory, expressed by the critical coupling strength (Eq. \eqref{kappac}), with 
simulations on the proposed dense small-world
networks, we use as topology functional the approximation for large systems (cf. Eq. \eqref{approxftop}).

In the following we further consider a Gaussian frequency distribution
$g_{\rm{gauss}}(\omega)$ with vanishing mean and standard deviation $\sigma$. Then
for the diversity functional \eqref{func_div} the expression
\begin{equation}
  f_{\rm{div}}(D)[g_{\rm{gauss}}]=2\sqrt{\frac{2}{\pi}}\sigma\left[1-\Phi\left(\frac{D}{\sqrt{2}\sigma}\right)\right]^{-1}\mathrm e^{-\frac{1}{2}\frac{D^2}{\sigma^2}}
\label{f_diff}
\end{equation}
follows; $\Phi(.)$ is the error function. The formula is equivalent to
the mean field expression derived in \cite{StrMir91} by means
of an eigenvalue analysis. In appendix \ref{fdiv_differentg} we summarize the 
limiting cases of Eq. \eqref{f_diff} and we give a comparison with other frequency distributions.

\begin{figure}
\begin{psfrags}
\psfragscanon
\psfrag{kappac}[t][t]{\color[rgb]{0,0,0}\setlength{\tabcolsep}{0pt}\begin{tabular}{c}$\varkappa_c$\end{tabular}}
\begin{centering}
\subfigure[$\ \alpha=0.05$]{\label{kappac_p}\includegraphics[width=0.49\linewidth]{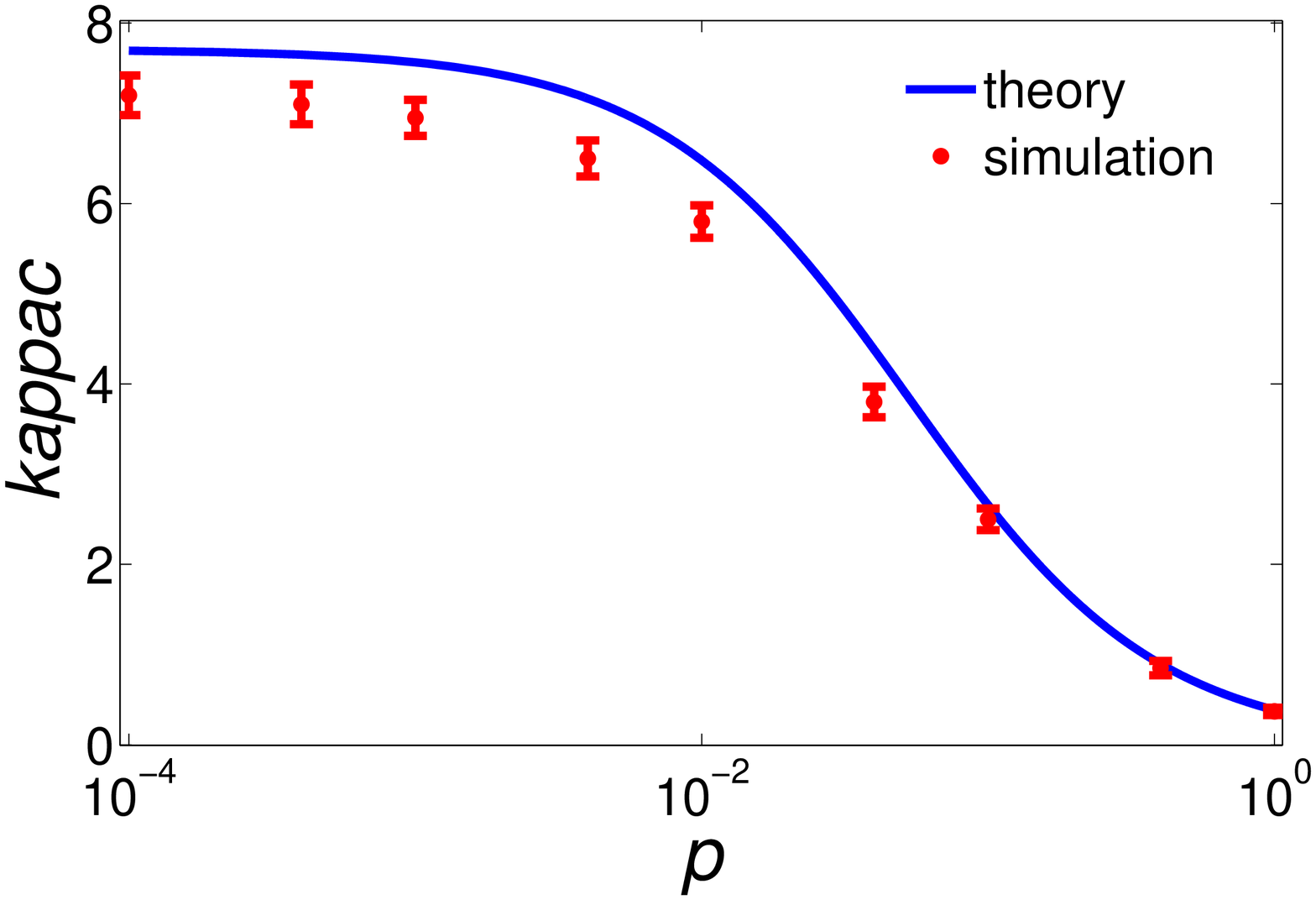}}
\subfigure[$\ p=0.01$]{\label{kappac_alpha}\includegraphics[width=0.49\linewidth]{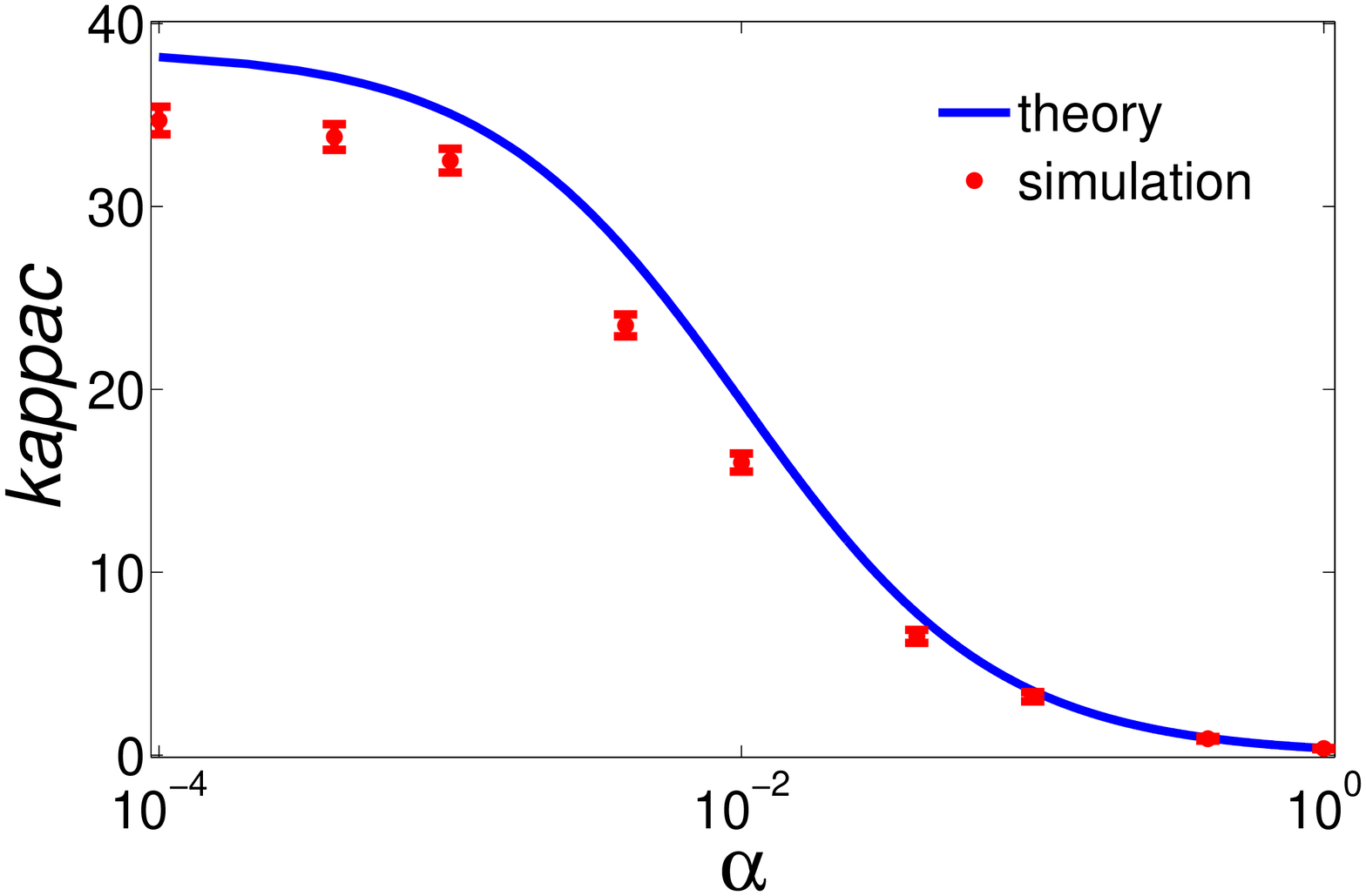}}
\caption[For $p>0.1$ or $\alpha>0.1$ we obtain almost a perfect agreement between theory and simulations.]{(Color online) The dependencies of 
the critical coupling strength $\varkappa_c$ on the shortcut probability $p$ and on the parameter $\alpha$ for the local connections
are shown. Remaining parameters: $D=0.05$, $\sigma=0.2$.}
\label{kappac_top}
\end{centering}
\end{psfrags}
\end{figure}

\begin{figure}
\begin{psfrags}
\psfragscanon
\psfrag{kappac}[t][t]{\color[rgb]{0,0,0}\setlength{\tabcolsep}{0pt}\begin{tabular}{c}$\varkappa_c$\end{tabular}}
\begin{centering}
\subfigure[$\ \sigma=0.2$]{\label{kappac_D}\includegraphics[width=0.49\linewidth]{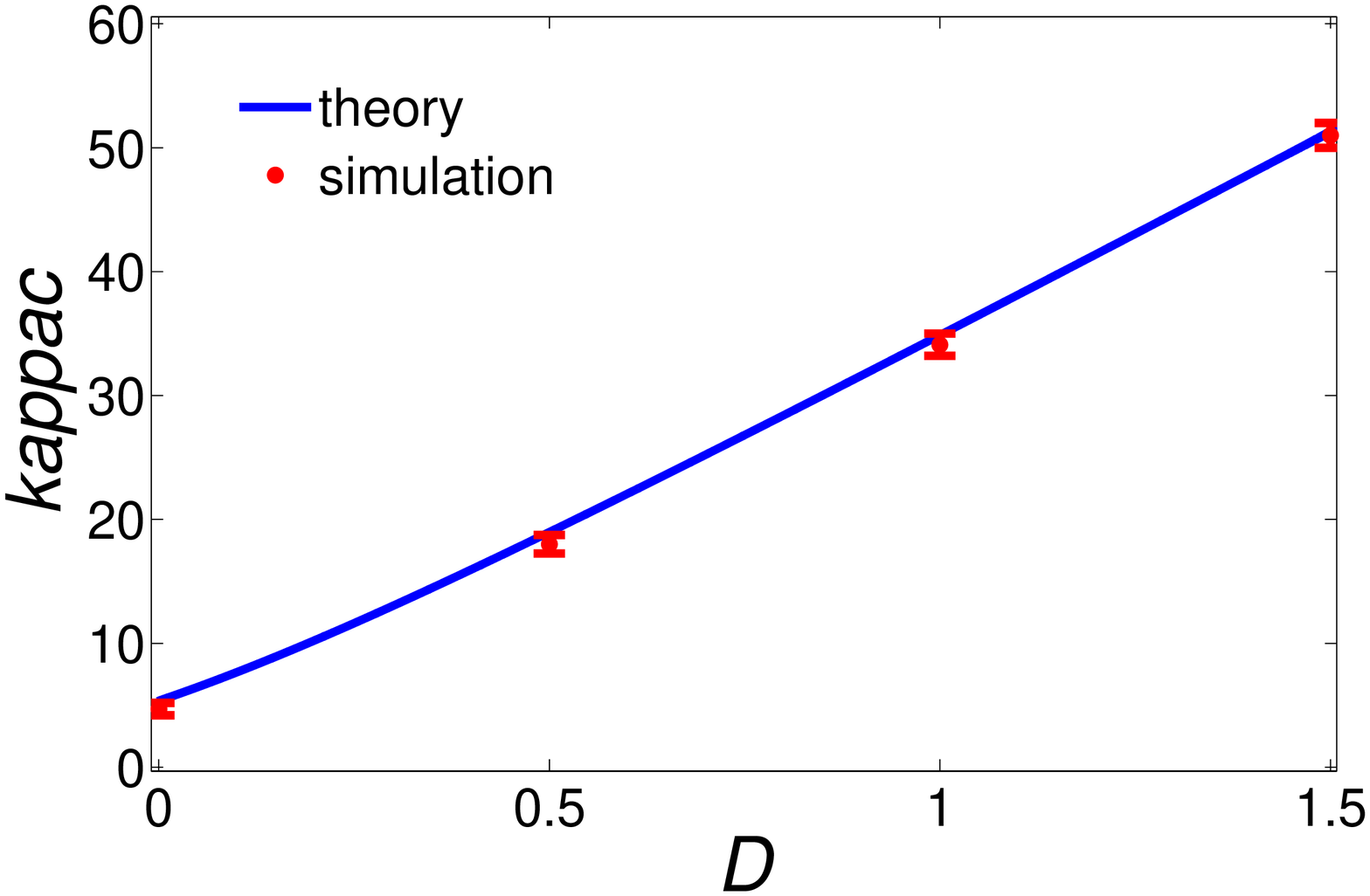}}
\subfigure[$\ D=0.05$]{\label{kappac_sigma}\includegraphics[width=0.49\linewidth]{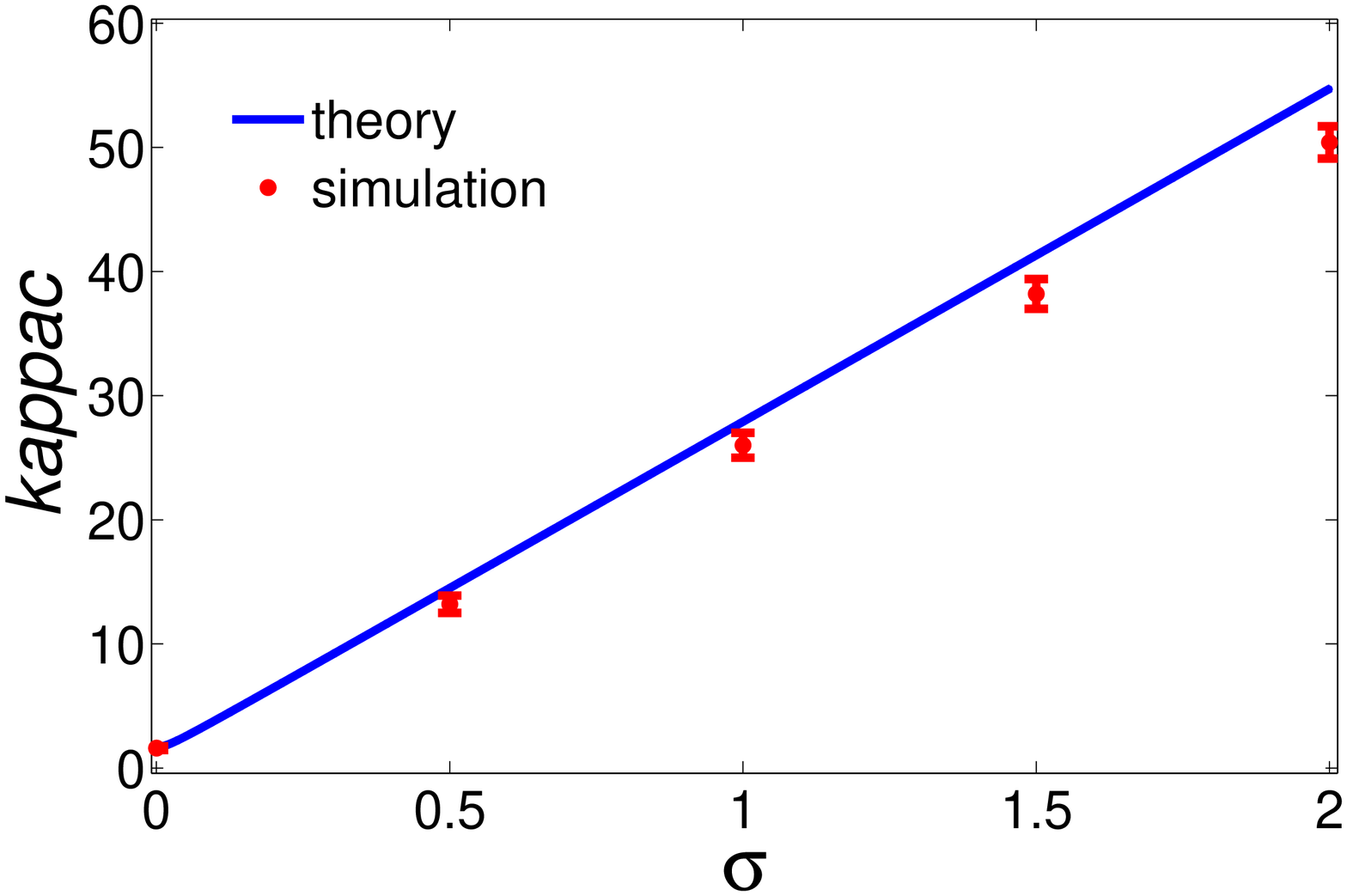}}
\caption[For the dependency on the noise intensity $D$ we obtain almost a perfect agreement between theory and simulations.]{(Color online) The dependencies of
the critical coupling strength $\varkappa_c$ on the noise intensity $D$ and on the diversity parameter $\sigma$ are depicted.
Remaining parameters: $\alpha=0.05$, $p=0.01$.}
\label{kappac_div}
\end{centering}
\end{psfrags}
\end{figure}

As can be seen in Figs. \ref{kappac_top} and \ref{kappac_div} we
obtain a satisfying agreement for the critical coupling strength
$\varkappa_c$, regardless of whether we consider the dependencies on
the topology (Fig. \ref{kappac_top}) or on the diversity (Fig. \ref{kappac_div}). 

Especially for $p>0.1$, $\alpha>0.1$ or $\sigma<0.5$ and for the dependency on the noise
intensity $D$ in general, we obtain almost a perfect agreement between theory and
simulations. For smaller values of $p$ or $\alpha$ there is a small
discrepancy, but the shape of the curves can be well reproduced.

As expected, both weaker connectivity and larger diversity impede synchronization. Compare Eq. \eqref{ftop_limiting} and Tab. \ref{fdiv_limiting} for a summary of the limiting cases. In particular, for $p\rightarrow 0$ or $\alpha\rightarrow 0$ our results suggest a saturation at the values $\varkappa_c=\frac{f_{\rm{div}}(D)[g]}{\alpha}$ or $\varkappa_c=\frac{f_{\rm{div}}(D)[g]}{p}$, respectively.

Apparently, the weighted mean-field approximation tends to overestimate the critical coupling strength,
which is counterintuitive, because the approximation corresponds to a weighted fully connected network
and all-to-all connectivity should reduce the critical coupling strength. To see this,
compare $p=1$ or $\alpha=1$ in Fig. \ref{kappac_top}, because both choices stand for all-to-all connectivity.
So the coupling weights defined in Eq. \eqref{newA} are able to mimic the original complexity in an overstated manner.

The disagreement for high standard deviations $\sigma$ of the Gaussian frequency distribution seems to be analogous to the disagreement in the dependency on the topology, because in our derivation of the critical coupling strength, both the frequency distribution $g(\omega)$ and the degree distribution $P(k)$ are involved in averaging.

\section{Beyond the Dense Small-World Networks}
\label{beyond}
\colr{Since the mean-field description corresponds to all-to-all connectivity, we expect
that the approximation loses validity in sparser networks.
In order to effectively analyze the limitations, we consider separately Erd\H os-R\'enyi like random networks and regular networks.
The first are constructed by assigning an edge probability of
\begin{equation}
p_e=p\cdot N^{q-1},\ 0\leq p,q\leq 1 
\label{pe}
\end{equation}
for any two of the $N$ nodes in the network. The only additional requirement is that besides the edge probability, each node connects a
priori to another randomly chosen one. In this way we guarantee that there are no isolated nodes, which are not of interest here,
because they are not able to take part in the synchronization process and only reduce the effective system size.}

\colr{In the regular networks each node is coupled to the
$K$ next neighbors in both directions of the ring network and $K$ is
given by (compare Eq. \eqref{Kalpha})
\begin{equation}
K=\left\lfloor\frac{1}{2}\left[3+\alpha\cdot(N-4)^q\right]\right\rfloor,\ 0\leq\alpha,q\leq 1.
\label{newKalpha}
\end{equation}
As for the random networks, we require that there are no isolated nodes, which is already implemented in the above choice ($K\geq 1$).}

\colr{In both cases $q$ is a denseness parameter, for $q=1$ leads to dense and $q=0$ to sparse networks \cite{Pr98}.}

\colr{We repeat the former analysis, see section \ref{comparison}, with adjusting the normalization of the coupling, i.e.,
we substitute $N\rightarrow N^{q}$ in front of the coupling term (cf. Eq. \eqref{ourmodel}). Consequently, the topology functional becomes
\begin{equation}
f_{\rm{top}}[P]=N^q \frac{\langle k\rangle}{\langle k^2\rangle}.
\label{newtop_func}
\end{equation}
In case of the regular networks it appears necessary to increase the system size for sparser networks in order to have distinguishable
networks (see Eq. \eqref{newKalpha}), e.g. for $\alpha=0.001$ and $q=0.75$ we consider network sizes up to $N=135000$. Again, $p=10^{-4}$ and
$\alpha=10^{-4}$ are the smallest values considered in our numerical simulations.}

\colr{In Fig. \ref{plot} the results are depicted. 
\begin{figure}
\centering
\includegraphics[width=0.49\linewidth]{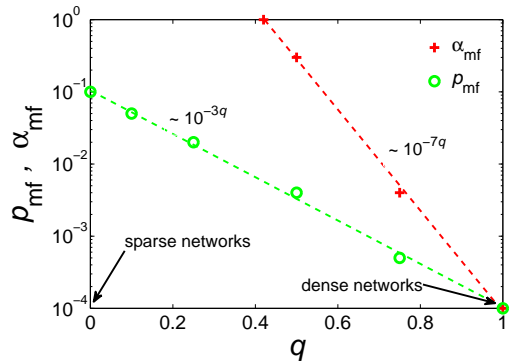}
\caption[]{\colr{(Color online) Smallest $p$ (random network) and $\alpha$ (regular network) values for which a synchronization
transition with the standard mean-field exponents can be observed, are shown as functions of the denseness parameter $q$.
We denote these values by $p_{\rm{mf}}$ and $\alpha_{\rm{mf}}$. Dashed lines depict the scaling behaviors.}}
\label{plot}
\end{figure}
Unlike below the lines, we observe a mean-field synchronization transition above them. In particular, the markers correspond with the smallest $p$ or $\alpha$ values for which we observe a synchronization transition with the critical mean-field exponents $\beta=\frac{1}{2}$ and $\nu=2$ \cite{HCK02}. Those $p$ and $\alpha$ values are denoted by $p_{\rm{mf}}$ or $\alpha_{\rm{mf}}$.}

\colr{As expected, for sparser networks the mean-field approximation breaks down. Furthermore, it can be seen that random shortcuts favor the
mean-field behavior, since we find the scaling relations $p_{\rm{mf}}\sim 10^{-3q}$ and $\alpha_{\rm{mf}}\sim 10^{-7q}$.}

\colr{We also underline that our approach works, if 
the first two moments of the degree distribution exist. So, generally,
scale-free networks that display a power-law decay $P(k)\sim k^{-\gamma}$
are excluded from our approximation. In \cite{ResOttHu05} it was
numerically shown that for $\gamma<3$ the mean-field approach yields
significant deviations. It is further analytically derived in
\cite{Lee05} that only for $\gamma>5$ the order parameter fulfills a
finite-size scaling with the standard mean-field exponenents, and for
$2<\gamma<3$ the exponents depend on the degree exponent $\gamma$. Hence,
more homogeneous degree distributions that can be characterized by the first two moments, favor the mean-field treatment as it was the case in our small-world example, see Fig. \ref{Pk}.}

\section{Conclusion}
\label{conclusion}
We have investigated noisy Kuramoto oscillators on undirected ring networks, whose complex structure is approximated by a weighted fully connected network.
The weights are obtained by a specific combinatorial consideration of the connectivity\colr{, where the original network is considered to be random and uncorrelated.
We} have shown that this procedure leads to a weighted mean-field approximation, which enabled us to evaluate analytically the critical coupling strength $\varkappa_c$
that marks the onset of synchronization in the network. As a result, we have found that $\varkappa_c$ is a product of two functionals. The first one is a functional
of the degree distribution and therefore depends solely on the network topology, while the second one is a functional of the frequency distribution and a function
of the noise intensity.

As such, we have provided support for the previous separate consideration of network complexity and noise with regard to the Kuramoto model in the literature.

In order to compare the theory with simulations, we have investigated a dense small-world network model. As typical of small-world networks \cite{NewWa99},
we start with a regular network consisting only of next neighbor couplings, and then we randomly add shortcuts. In this way one can study the impact and
the interplay of regular local edges and random shortcuts. We have found that in a large small-world network, regular local edges and random shortcuts play almost the same role.

Our dense small-world network model allows scaling between a locally and a globally connected network, which is well-suited for testing the mean-field approximation.
In case of all-to-all connectivity the mean-field description is exact; by decreasing the connectivity we can determine the arising discrepancies between the original complexity and our estimate. 

It has turned out that the analytical and the numerical results are consistent to each other in particular, finite-size scaling analysis with mean-field exponents always gives a well-defined
intersection point. Hence, the extended Kuramoto model on dense small-world networks shows a mean-field synchronization transition.

\colr{We have completed our work by addressing further network models. There we have shown that random connections favor a mean-field approach in the sense that regular networks have to be more
densely connected in order to yield a mean-field synchronization transition. Highly heterogeneous networks in which the mean-field description breaks down, have also been discussed.}

Future work has to be done with regard to network complexity. For instance whether networks of higher dimension show significant qualitative differences. In particular, the problem of
time-dependent coupling strengths or even time-dependent number of nodes and edges has to be tackled. A next step could also be the consideration of time delays in the coupling.
Furthermore, network costs have to be taken into account, which should allow to evaluate the efficiency of a synchronization process.

\acknowledgments
This work was supported by GRK1589/1. \colr{We acknowledge \mbox{F. Sagu\'es}, \mbox{S. Martens} and \mbox{P. K. Radtke} for a critical reading of the manuscript and useful comments.}

\appendix
\section{Dense Small-World Networks}
\label{dense_sw}
\begin{figure}
\begin{centering}
\subfigure[$\ \alpha=0.008$]{\label{candl_fordense_0008}\includegraphics[width=0.49\linewidth]{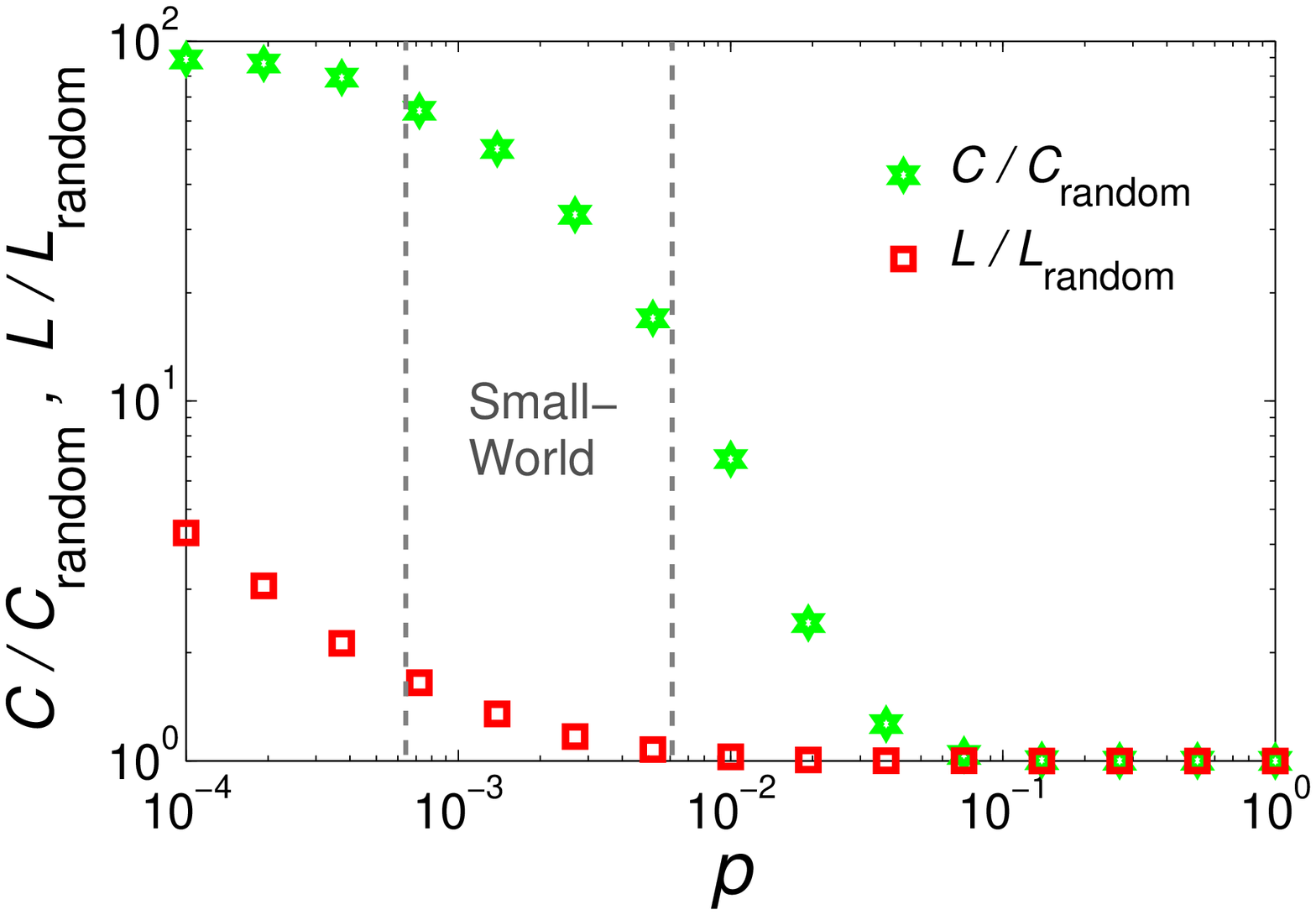}}
\subfigure[$\ \alpha=0.2$]{\label{candl_fordense_02}\includegraphics[width=0.49\linewidth]{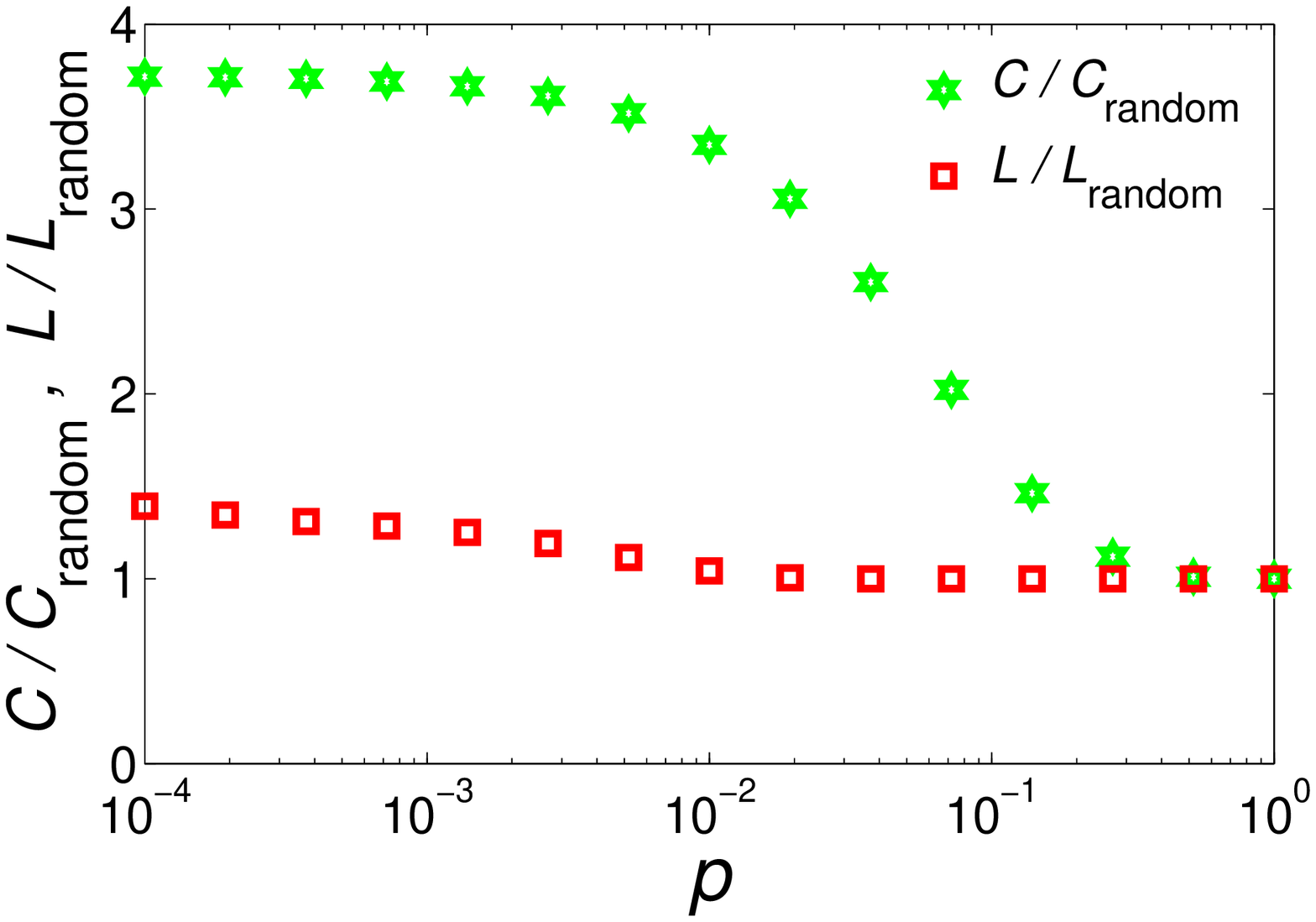}}
\caption[Numerical evidence for dense small-world networks.]{(Color online) Numerically calculated average
path lengths $L$ and clustering coefficients $C$ as a function of shortcut probability $p$ for a network of 501 nodes. 
The radius of coupled next neighbors (compare Eq. \eqref{Kalpha}) equals $K=2$ (Fig. \ref{candl_fordense_0008})
or $K=50$ (Fig. \ref{candl_fordense_02}), respectively. Dashed lines indicate the area, in which our model generates a small-world network.}
\label{candl_fordense}
\end{centering}
\end{figure}

Here we provide numerical evidence for the existence of dense small-world networks, see Fig. \ref{candl_fordense}. In order to do this, we have to compare such dense small-world networks with random networks. However, the random network used for the comparison must not have any isolated nodes. Otherwise path length and clustering coefficient are not (well) defined. We take as an almost random network (similar to Watts and Strogatz \cite{WaStr98}) a network generated with our model, but with a minimum amount of coupled next neighbors, so we choose $K=1$ or an equivalent $\alpha$ (cf. Eq. \eqref{Kalpha}). The comparison makes sense, only if the average degree $\langle k\rangle$ of both networks is the same, which requires choosing the following shortcut probability $p'$ for the random network:
\begin{equation}
p'=\frac{2(K-1)}{N-3}+\frac{N-2K-1}{N-3}p\ .
\end{equation}
$p$ is the shortcut probability used for the dense small-world network.

Having in mind the defining properties of a small-world network: $L\gtrsim L_{\mathrm{random}}$ and $C\gg C_{\mathrm{random}}$ \cite{WaStr98}, we observe small-world networks for values of $p$ around $10^{-3}$. For smaller values of $\alpha$, i.e., fewer coupled next neighbors, the defining properties are better fulfilled. If $\alpha$ is too large the network under consideration is no small-world network anymore.

\section{Diversity Functional for Different Frequency Distributions}
\label{fdiv_differentg}
In what follows, we consider the diversity functional for different frequency distributions: uniform distribution $g_{\rm{uni}}(\omega)$, identical oscillators $g_{\rm{ident}}(\omega)$ and Lorentzian distribution $g_{\rm{lorentz}}(\omega)$. We obtain the following expressions for the diversity functionals (cf. Eq. \eqref{func_div}):
\begin{equation}
\begin{aligned}
f_{\rm{div}}(D)[g_{\rm{uni}}]&=2\sqrt{3}\sigma\left[\arctan\left(\frac{\sqrt{3}\sigma}{D}\right)\right]^{-1},\\
f_{\rm{div}}(D)[g_{\rm{ident}}]&=2D,\\
f_{\rm{div}}(D)[g_{\rm{lorentz}}]&=2(D+\gamma).
\end{aligned}
\end{equation}

\begin{figure}
\begin{centering}
\subfigure[$\ \sigma=\gamma=1$]{\label{fdiv_D}\includegraphics[width=0.49\linewidth]{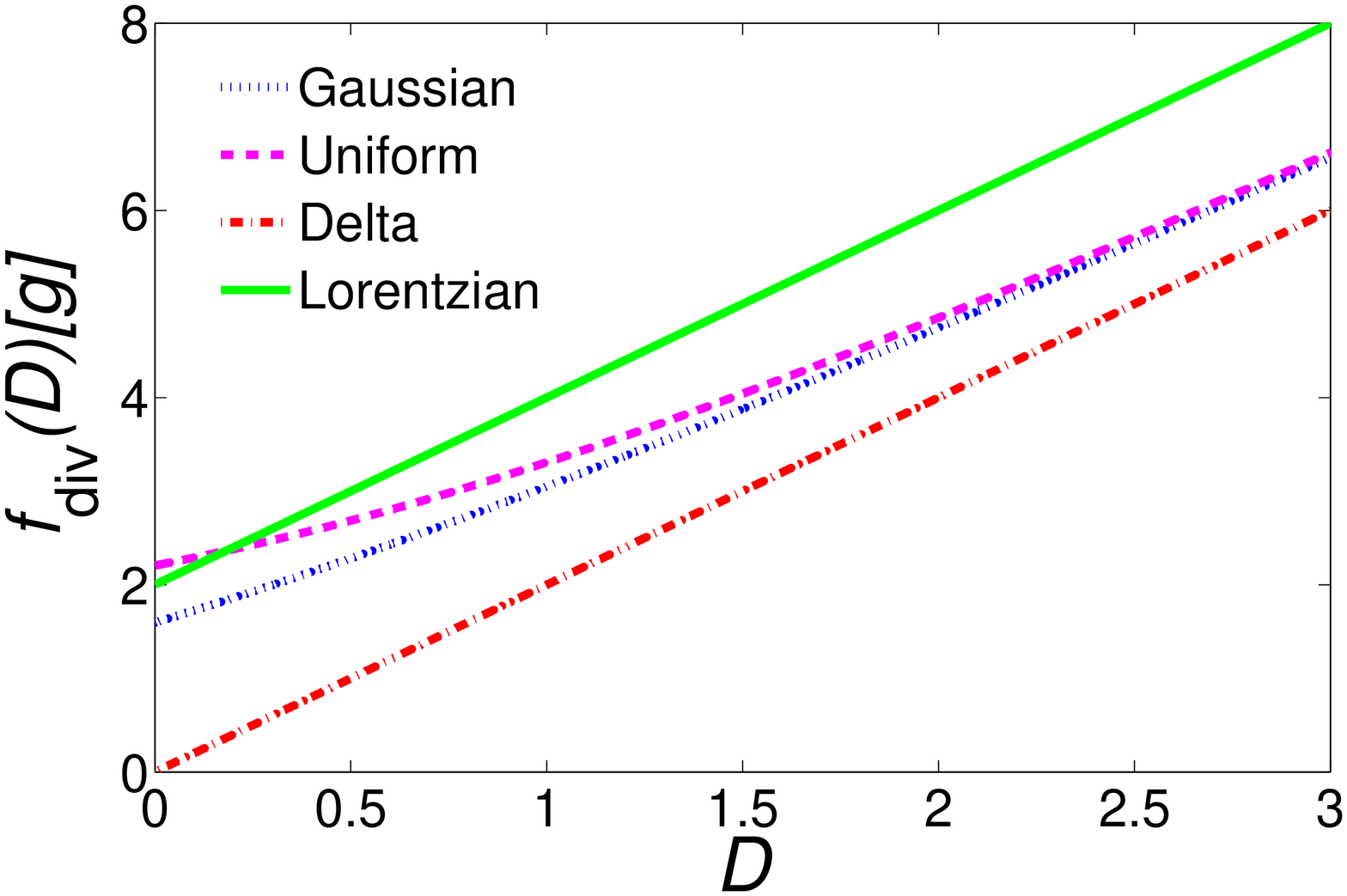}}
\subfigure[$\ D=0.4$]{\label{fdiv_sigma}\includegraphics[width=0.49\linewidth]{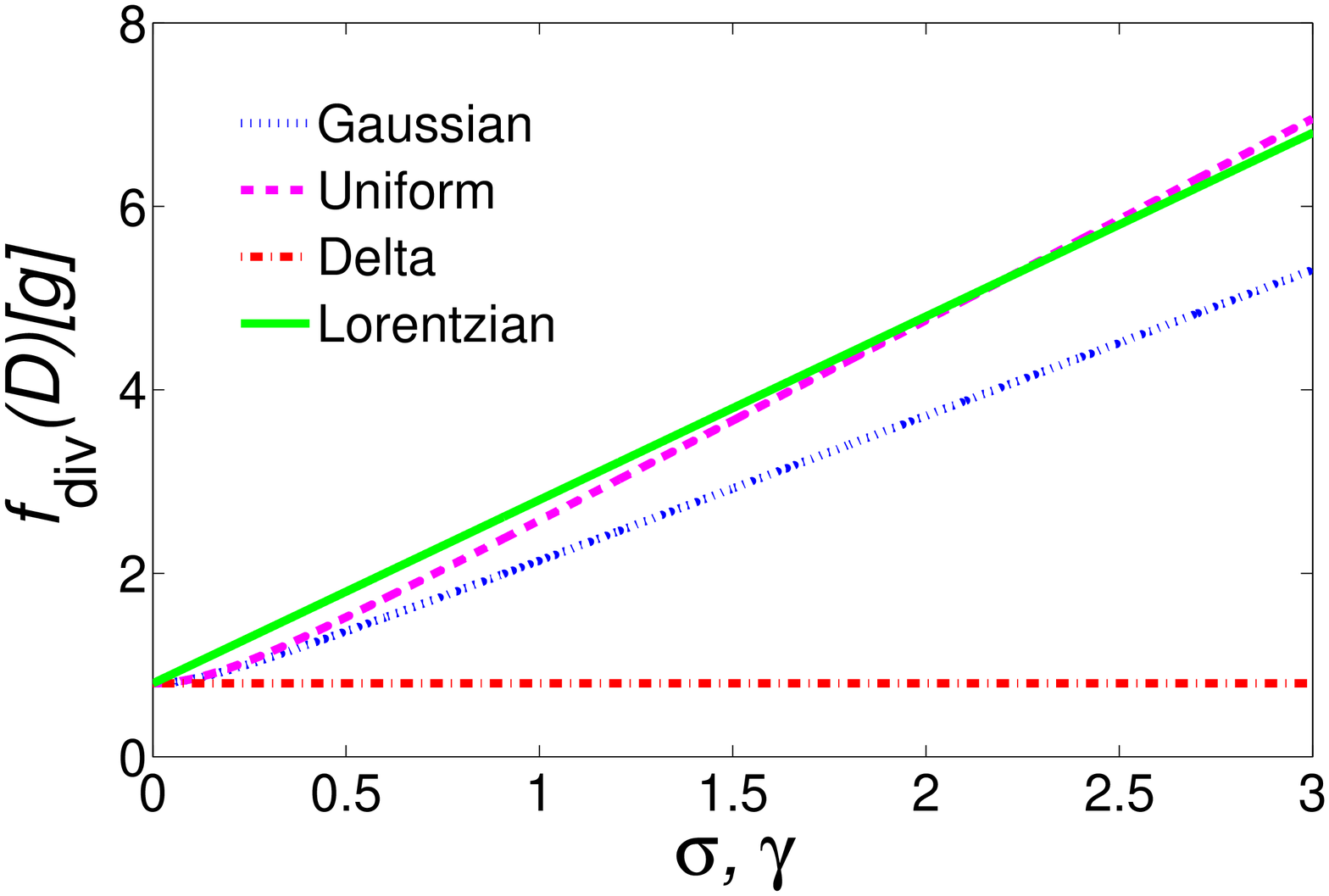}}
\caption[We always find linear dependencies of the diversity functional for large arguments.]{ (Color online)
The diversity functional for different frequency distributions as a function of noise intensity $D$,
standard deviation $\sigma$ or scale parameter $\gamma$.}
\label{fdiv}
\end{centering}
\end{figure}

$\sigma$ and $\gamma$ are the standard deviation and the scale parameter, respectively, while $D$ gives the noise intensity.
In Fig. \ref{fdiv} the dependencies on $D$, $\sigma$ or $\gamma$ are depicted. We observe a greater difference in the dependency on diversity than on noise. In particular, for large values of $D$, all the different diversity functionals show the same linear dependency on $D$ (cf. Fig. \ref{fdiv_D}). In case of Gaussian, uniformly and identically distributed natural frequencies $\omega$, it can be shown that the diversity functional even approaches the same line for $D\rightarrow\infty$.

However, the above observation is not surprising, because the noise acts on the natural frequencies. So if the noise intensity $D$ is very high, it makes almost no difference how the natural frequencies are distributed, because the diversity of the oscillators mainly comes from the random fluctuations induced by the noise.

Instead, for $D=\rm{const.}$ and increasing diversity parameter $\sigma$ or $\gamma$, the different nature of the various frequency distributions manifests more and more in a different synchronizability (cf. Fig. \ref{fdiv_sigma}). Especially the comparison between the uniform and the Lorentzian distribution for the same $\sigma$ and $\gamma$ values is interesting. In terms of synchronizability we observe that for lower diversity the uniform frequency distribution is more 
favorable than the Lorentzian distribution, while for higher diversity the Lorentzian distribution is more favorable.

Of course, vanishing diversity, i.e., $\sigma\rightarrow 0$ or $\gamma\rightarrow 0$, results in identical oscillators, so that the diversity functional approaches the same value for every frequency distribution. In Tab. \ref{fdiv_limiting} we summarize these results.

\begin{table}[htp]
\begin{center}
\begin{tabular}{c|c|c|c|c}
\toprule
\textsc{limiting case} & $D\ll1$ & $\sigma\ll1$ & $D\gg1$ & $\sigma\gg1$ \\
\midrule
$f_{\rm{div}}(D)[g_{\rm{gauss}}]$  &  $2\sqrt{\frac{2}{\pi}}\sigma+\frac{4}{\pi}D+O(D^2)$ & $2D+\frac{2}{D}\sigma^2+O(\sigma^4)$ & $2D+O\left(\frac{1}{D}\right)$ & $2\sqrt{\frac{2}{\pi}}\sigma+O(1)$ \\
$f_{\rm{div}}(D)[g_{\rm{uni}}]$ & $\frac{4\sqrt{3}}{\pi}\sigma+\frac{8}{\pi^2}D+O(D^2)$ & $2D+\frac{2}{D}\sigma^2+O(\sigma^4)$ & $2D+O(\frac{1}{D})$ & $\frac{4\sqrt{3}}{\pi}\sigma+O(1)$ \\
$f_{\rm{div}}(D)[g_{\rm{ident}}]$ & $2D$ & $2D$ & $2D$ & $2D$ \\
$f_{\rm{div}}(D)[g_{\rm{lorentz}}]$ & $2\sigma+2D$ & $2D+2\sigma$ & $2D+O(1)$ & $2\sigma+O(1)$ \\
\bottomrule
\end{tabular}
\caption{The limiting cases of the diversity functional for different frequency distributions. Here $\sigma=\lambda$.}
\label{fdiv_limiting}
\end{center}
\end{table}

\bibliography{bibliography}
\bibliographystyle{unsrt}
\end{document}